\documentclass[fleqn]{2017SCGE}

\usepackage{amsmath}
\usepackage{amssymb}

\begin{document}

\ensubject{subject}




\title{Theoretical study on bubble dynamics under hybrid-boundary and multi-bubble conditions using the unified equation}{Theoretical study on bubble dynamics under hybrid-boundary and multi-bubble conditions using the unified equation}
\author[1,2]{A-Man Zhang}{{zhangaman@hrbeu.edu.cn}}%
\author[1]{Shi-Min Li}{}
\author[1,2]{Pu Cui}{}%
\author[1,2]{Shuai Li}{}%
\author[1,2]{Yun-Long Liu}{}

\AuthorMark{A-Man Zhang}

\AuthorCitation{Zhang A-Man, Li Shi-Min, Cui Pu, et al}

\address[1]{College of Shipbuilding Engineering, Harbin Engineering University, Harbin, 150001, China}
\address[2]{Nanhai Institute of Harbin Engineering University, Sanya, 572024, China}


\abstract{This paper aims to use the unified bubble dynamics equation to investigate bubble behavior in complex scenarios involving hybrid free surface/wall boundaries and interactions between multiple bubbles. The effect of singularity movement on the unified equation's form is analyzed after deriving the bubble pulsation equation using a moving point source and a dipole, followed by discussions on the effect of migration compressibility-related terms on the bubble dynamics. In addition, the present study accounts for the impact of hybrid boundaries, including crossed and parallel boundaries, by introducing a finite number of mirror bubbles for the former and an infinite number of mirror bubbles for the latter. Spark bubble experiments and numerical simulation are conducted to validate the present theory. The application of the unified equation in multi-bubble interactions is exemplified by computing a spherical bubble array containing more than 100 uniformly distributed cavitation bubbles under different boundary conditions. The bubble cluster-induced pressure peak can reach nearly two times or even higher than that of an individual bubble, highlighting the damage potential caused by cavitation bubble clusters.}



\maketitle


\begin{multicols}{2}
\section{\label{sec:intro}Introduction}
Bubble dynamics find extensive application in scientific and engineering fields, encompassing marine science and engineering \cite{ loewen1996bubbles, Deike2022, Watson19}, advanced manufacturing \cite{Nourhani2020,Lohse2022,turkoz2018impulsively,ji2016numerical}, environmental and chemical engineering \cite{spzh14,carling2017bubble,Yang2022}, medicine \cite{sc19,Dollet2019,Rod2015,Mukundakrishnan2009}, and life sciences \cite{vas12, Lohse2018D,vshl00}. For instance, large-scale underwater explosion bubbles \cite{Cole1948book,Cui2016,zhang2021engineering,lhfww19} can generate massive spike/tsunamis, high-pressure air gun bubbles \cite{Ziolkowski1982,de2014bubble,wehner2020acoustic} are effective pressure pulse sources in deep sea exploration, cavitation bubbles are key factors in water exit/entry of objects \cite{Aristoff2009,Truscott2014,Wentao2023}, and controllable small-scale bubbles \cite{Brennen2015,lhgh18,omata2022ultrasound} can be important carriers in life science fields. 
Therefore, bubble dynamics is an essential scientific problem in many fields. 
In a practical environment, bubbles often exist in swarms or face complex boundary conditions. This could cause unpredictable changes in bubble dynamics. An in-depth understanding of the multi-bubble coupling or  bubble-boundary interaction is vital for using the bubble energy efficiently or avoiding the hazardous effects caused by bubbles. 

As early as 1917 and 1949, Rayleigh \cite{r17} and Plesset \cite{plesset49} established the bubble pulsation equation in a incompressible flow field. In 1970, Hicks \cite{h70} modified the Rayleigh-Plesset equation by considering incompressible migration. Keller et al.\cite{km80} derived the bubble pulsation equation in compressible flow fields based on the wave equation in 1980. Chahine \cite{Chahine2009} included the bubble migration in Keller equation by adding an incompressible migration term in 2009. These theories have played an important role in the bubble dynamics field. However, because bubbles are typically not isolated, their dynamic behavior is obviously affected by boundary effects, multiple bubbles, flow field environment, gravity, and many other conditions. These factors can cause bubbles to have pulsation and compressible migration. However, the above-mentioned theories did not fully consider bubble pulsation and migration in a compressible flow field. 

To more accurately predict bubble dynamics in theory, Zhang et al. \cite{zhang2023unified} constructed the basic solution of the moving singularity of the wave equation from the most basic mathematical principles while considering the compressible migration effect of bubbles in 2023. Through a series of mathematical and physical calculations and experimental verification, they derived a unified equation for bubble dynamics with an elegant mathematical form that can simultaneously consider boundary effects, multiple bubbles, flow field environment, gravity, migration, compressibility, viscous force, and surface tension. The unified equation enables a theoretically accurate prediction of bubble dynamic behavior, guiding experiments and predictions of new physical phenomena.

Extensive studies on the dynamics of single bubbles near a single boundary, including the rigid wall \cite{Gonzalez2021, Blakewall1986, qianxi2014}, free surface \cite{saade2021,Koukouvinis2016,Koukouvinis2016}, and elastic boundary \cite{Goh2017, tokk06,Brujan2001}, have uncovered and clarified many interesting physical phenomena: re-entrant jet, annular jet, and water spike on the free surface. In recent years, bubble dynamics under the action of multiple bubbles or multiple boundaries have also received much attention. Two bubbles generated simultaneously produce two liquid jets in opposite directions \cite{Han2015,Tong2019}, while the collapse patterns of bubbles generated unsynchronously are highly correlated with the phase difference \cite{liang2021interaction,Tomita2017s}. Some special phenomena, such as the mutual penetration between bubbles and fine thin jets, are found by Liang et al.\cite{liang2021interaction} and Tomita et al.\cite{Tomita2017s}. When the bubble buoyancy effect is considered, the bubble jet direction is deflected upward \cite{liu2021interaction, Han2019s}, altering the bubble migration direction. When the third bubble is included, it enhances or weakens the strength of the bubble jets generated by the other two bubbles depending on the phase and size difference of the three bubbles \cite{Chen2022}. In addition, for bubble clusters formed by more bubbles, the bubbles inside the cluster are accelerated in collapsing by the surrounding bubbles and generate a much stronger pressure pulse than that of a single bubble \cite{Zwaan2007,amzhang2023,zhang2019numerical}. Different ambient flow fields could significantly alter the dynamic characteristics of bubble clusters \cite{Doinikov07, Maeda2019}. For bubble dynamics near multiple boundaries, when two rigid walls affect the bubble dynamics in different directions, the bubble jet is directed toward the closer wall \cite{RN213,Qianxi20c,CuiRN2020}. Furthermore, when the bubble buoyancy effect is considered, the jet may turn from downward to upward upon improving the buoyancy effect \cite{Shimin21c}. When both a free surface and a sidewall are present at similar distances to the bubble, the bubble is more affected by the free surface, producing a downward but slightly inclined jet \cite{Yun2016c,ZhangS2017c,Kiyama2021,Shim2023v}. \cite{shimin2019} investigated three boundaries, including the free surface, sidewall, and bottom wall, and linked the various bubble collapse patterns (e.g., jetting and quasi-spherical collapsing) to distance parameters and the buoyancy effect.

The above studies used numerical simulations and experiments to investigate bubble dynamics in different environments in most cases. Numerical methods such as boundary element \cite{Blake87b,Zhangm2013,Wang03BIM, Hanrui2021}, finite element \cite{Liu19i,LiuNN2020,Wang16d}, finite volume \cite{Zeng20c,Zeng18c,Lechner20c}, and smooth particle hydrodynamics methods \cite{Zhang15ph,li2022improved} have been greatly developed in the past decades for predicting bubble behavior and revealing its underlying mechanism. Experimental methods such as laser pulse \cite{RN112021, altmvl01, ro21, Jingzhu2021}, electric discharge \cite{Dadvand2012,Goh2017,okdbnf06,sun2022cavitation} and small charge underwater explosions \cite{Cui2016, hh10, Shimin21d} are mature. However, when handling farfield problems or a large number of bubbles, both simulations and experiments can be costly. In such cases, theoretical models prove superior for investigating the volume oscillation, migration, and field pressure of a large number of bubbles. 

In this study, we derive a unified equation using a moving point source and dipole, and the effects of singularity motion and compressible bubble migration are discussed. We then further apply the theory to bubble dynamics near a hybrid boundary, including crossed and parallel boundaries, and compare the results with experiments and numerical simulations. The dynamics of three-dimensional bubble arrays are also discussed by computing a spherical bubble cluster containing over 100 cavitation bubbles under different boundary conditions, using the derived equation. Computed results show that the bubbles in the cluster could have greater damage potential to the surrounding structure than a single bubble. 

\section{Derivation of the unified equation for bubble dynamics}\label{sect:basic-theory}
\subsection { Point source and dipole for moving bubble}

In the compressible flow, the conservation equations for mass and momentum are
\begin{equation}
\label{eq1}
\frac{{\partial \rho }}{{\partial t}}  = - \nabla  \cdot \left( {\rho \boldsymbol{u}} \right)
\end{equation}	

\noindent and 

\begin{equation}
\label{eq2}
\frac{ \partial (\rho \boldsymbol{u})}{\partial t} + \nabla  \cdot \left( {\rho \boldsymbol{u} \boldsymbol{u}} \right)
= -\nabla p + \nabla \cdot \boldsymbol{\tau} +  \rho \mathcal{F},
\end{equation}
where $\rho$, $\boldsymbol{u}$, $p$, and $\boldsymbol{\tau}$ represent the pressure, density, velocity, and deviatoric stress tensor of the fluid, respectively; $\mathcal{F}$ is the body force. 

Following the previous works, we assumed that the fluid around the bubble is weakly compressible \cite{zhang2023unified}, and the wave equation is obtained:
\begin{equation}
\label{eq-wave-equation}
\frac{1}{{{C^2}}}\frac{{{\partial ^2}\varphi }}{{\partial {t^2}}}   = {\nabla^2}\varphi,
\end{equation}
in which $C$ is sound speed in water and $\varphi$ is the velocity potential.

In this study, the origin of a spherical coordinate system $o - r\theta \phi$ is set to the initial bubble center. Assume that the bubble remains spherical during the whole process.
$\theta=0$ denotes the bubble migration direction. In the previous study \cite{zhang2023unified}, we obtained a unified bubble equation by superimposing the velocity potential induced by a point source and a diple. In this study, we further derive the unified bubble equation by considering the movement of the two singularities. First, a source is set at the origin with the strength of $f_0(t)$, and the basic solution of the wave equation can be written as follows:
\begin{equation} \varphi_{f_0}(\boldsymbol{r},t) = \frac{1}{|\boldsymbol{r}|} f_0(t-\frac{|\boldsymbol{r}|}{C}).
\end{equation} 

As the point source and dipole move with an arbitrary velocity of $\bm v(t)$, the basic solution can be obtained by the superposition of infinite $\phi_{f_0}$ as follows:	
\begin{equation}\label{eq-source-moving}
{\color{black}\varphi_f(\boldsymbol{r},t) = 
	\frac{C}{|\boldsymbol{r}_t| \left(C-\boldsymbol{v}\cdot\boldsymbol{r}_t\right)} f(t-\frac{|\boldsymbol{r}_t|}{C})},
\end{equation}
where $\boldsymbol{r}_t$ denotes the vector pointing from the source at $t-|\boldsymbol{r}_t|/C$ to $\boldsymbol{r}$. 

\begin{figure}[H]
	\centering
	\includegraphics[width=7cm]{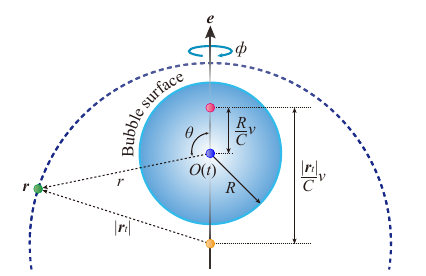}
	\caption{Geometric relations between an arbitrary location $\boldsymbol{r}$ (marked by the green dot), the moving singularities, and the spherical bubble. The blue dot denotes the origin of the coordinate system $o-r\theta\phi$ at time $t$. The singularities at $t - R/C$ coincides with the blue dot. The red and orange dots represent the locations of the singularities at $t$ and $t - |\boldsymbol{r}_t|/C$, respectively. }
	\label{fig:wave-eq-sketch}
\end{figure}

However, providing the analytic solution of ${\bm r}_t$ is challenging because a nonlinear equation is involved. With reasonable simplification, the migration velocity $v$ can be treated as a constant as the singularity propagates and exerts influence from the bubble center to the bubble surface. During this period, the displacement of the singularities $Rv/C$ is a small quantity relative to the bubble radius $R$. Figure \ref{fig:wave-eq-sketch} shows a sketch of the singularities and the propagation of their influence. $R$ denotes the bubble radius at the present time $t$. The position of the singularities at the time $t- R/C$ is the same as that of the origin at the time $t$ (blue dot). Thus, the singularity-emitted influences at $t- R/C$ reach the bubble surface ($r = R$) at $t$. 
Therefore, the distance from the present location of the singularities (red dot) to the origin $O(t)$ is $Rv/C$.
In addition, the velocity potential at $\boldsymbol{r}$ is determined by the singularity-emitted influences at time $t - |\boldsymbol{r}_t|/C$. The singularity position at that time is indicated by the orange dot. The distance between the location of the singularities at the present time $t$ and that at $t - |\boldsymbol{r}_t|/C$ can be easily obtained as $|\boldsymbol{r}_t|v/C$. Thus, we have the equation with a single unknown, $|\boldsymbol{r}_t|$, as follows:

\begin{equation}
|\boldsymbol{r}_t|^2 = \left(r\cos\theta-(t-t_0)v + (|\boldsymbol{r}_t|-R)\varepsilon\right)^2+\left(r\sin\theta\right)^2,
\end{equation}
where $\varepsilon=v/C$; $t_0$ is a reference time to offset the origin to the bubble center such that $t - t_0 = 0$. However, when calculating the derivative of $d$ and $\mathcal{D}$, $t - t_0$ should be retained. Then, $|\boldsymbol{r}_t|$ can be obtained by solving the above equation:

\begin{equation}\label{eq-express-d}
|\boldsymbol{r}_t| = \frac{1}{{1 - {\varepsilon ^2}}}\left( \begin{array}{l}
\cos \theta r\varepsilon  - tv\varepsilon  + {t_0}v\varepsilon  + \\
\sqrt {\begin{array}{l}
	{\cos ^2}\theta {r^2}{\varepsilon ^2} - 2r\cos \theta vt + 2r\cos \theta v{t_0}\\
	- {r^2}{\varepsilon ^2} + {v^2}{t^2} - 2{v^2}t{t_0} + {v^2}{t_0}^2 + {r^2}
	\end{array} }
\end{array} \right).
\end{equation}

Substituting equation (\ref{eq-express-d}) into equation (\ref{eq-source-moving}), we can obtain the expression of the moving source:

\begin{equation}
\label{movingsource}
\varphi_s {(r,\theta,t)} = \frac{1}{\mathcal{D}}f{\left(t-\frac{|\boldsymbol{r}_t|}{C}\right)},
\end{equation}
where $\mathcal{D}$ is expressed as follows:
\begin{equation}
\mathcal{D} = \sqrt{\left(r\cos\theta-(t-t_0)v - R\varepsilon \right)^2 + \left(r\sin\theta\right)^2 (1-\varepsilon^2)}.
\end{equation}

For a moving dipole with a strength of $q(t)$, the wave equation can be solved by taking the limit as follows:
\begin{equation}\label{eq-dipole-moving}
\varphi_q = \lim_{S \rightarrow0}\frac{1}{2S}\left(\varphi_f(\boldsymbol{r}+\boldsymbol{e}S,t)-\varphi_f(\boldsymbol{r}-\boldsymbol{e}S,t)\right),
\end{equation}
where ${\boldsymbol{e}}$ denotes a unit vector along the migration direction of the bubble.

Thus, the solution of the wave equation for a source with the strength $f$ and a diple with the strength $q$ both moving with a constant velocity $v$ can be expressed as follows:

\begin{equation}
\label{eq-velocity-potential}
\varphi_{(r,\theta,t)} = \frac{1}{\mathcal{D}}f{\left(t-\frac{|\boldsymbol{r}_t|}{C}\right)} + \frac{{Q}}{\mathcal{D}^3}q{\left(t-\frac{|\boldsymbol{r}_t|}{C}\right)} + 
\frac{\varepsilon}{v}\frac{{Q}+\mathcal{D}\varepsilon}{\mathcal{D}^2(1 - \varepsilon^2)}q'{\left(t-\frac{|\boldsymbol{r}_t|}{C}\right)},
\end{equation}
where $Q$ is written as 
\begin{equation}
{Q} = r\cos\theta-(t-t_0)v-R\varepsilon.
\end{equation}

Verifying that equation (\ref{eq-velocity-potential}) satisfies the wave equation is a straightforward task. Next, the velocity components in $r$ and $\theta$ directions at the bubble wall can be expressed as follows:
\begin{multline}
\label{eq-phi-r}
u_{r}{(R,\theta,t)} = \frac{\partial \varphi}{\partial r}  = 
- \frac{f}{R^2}- 
\frac{1}{RC} f' 
-\frac{2\cos\theta}{R^3}q \\- \frac{{9{{\cos }^2}\theta qv + 2R\left( {fv + q'} \right)\cos \theta  - 3qv}}{{C{R^3}}},
\end{multline}
and
\begin{multline}
u_{\theta}{(R,\theta,t)} = 
\frac{1}{R}\frac{\partial \varphi}{\partial \theta} \\=
- \frac{{\sin \theta }}{{{R^3}}}q - \frac{{\sin \theta \left( {6\cos \theta qv + R\left( {fv + q'} \right)} \right)}}{{C{R^3}}},
\end{multline}
respectively, where $f'$ and $q'$ denote the derivative of $f$ and $q$ with respect to their arguments, respectively. 

\subsection {Single bubble dynamics equation}

The kinetic boundary condition \cite{zhang2023unified} for oscillation and migration on the bubble surface can be expressed as follows:
\begin{equation}
\label{eq-kinetic-condition-osc}
\int_{\partial V} u_r \mathrm{d}S = 4\pi R^2\dot{R}
\end{equation}
and
\begin{equation}\label{eq-kinetic-condition-mig}
\frac{\mathrm{d}}{\mathrm{d}t}\int_V r\cos\theta \mathrm{d}V = \frac43\pi R^3v,
\end{equation}
respectively, where $V$ denotes the bubble volume and $\partial V$ denotes the bubble surface. $\dot{R}$ is the time derivative of the bubble radius. 
{\color{black}Denoting $F(t) = -f(t-R/C)$ and $Q(t) = q(t-R/C)$, and substituting equation (\ref{eq-phi-r}) into above two equations, we have} 
\begin{equation}
\label{eq-ODE-boundarycondition-osc}
F + \frac{R}{C-\dot{R}}\frac{\mathrm{d}F}{\mathrm{d}t} = R^2\dot{ R}
\end{equation}
and
\begin{equation}
\label{eq-ODE-boundarycondition-mig}
\frac{Q}{R} + \frac{{\dot Q}}{{C - \dot R}} + \frac{{Fv}}{C} =  - \frac{{R{^2}v}}{2},
\end{equation}
respectively.
 Using a perturbation method on equation (\ref{eq-ODE-boundarycondition-mig}) with perturbation parameter of $1/C$, we obtain the expression of $Q$ 
\begin{equation}\label{eq-s}
Q = - \frac{{vR{^3}}}{2} + \frac{1}{{2C}}\left( {5v{R^3}\dot R + {R^4}\dot v} \right),
\end{equation}
with only the terms up to the order $C^{-1}$ reserved. Differentiating both sides of equation (\ref{eq-ODE-boundarycondition-osc}) with respect to time, we have
\begin{equation}\label{eq-core-2}
\left( {\frac{{C - \dot R}}{R} + \frac{\mathrm{d}}{{\mathrm{d}t}}} \right)\left( \frac{{R}}{C - \dot R}\frac{\mathrm{d}F}{\mathrm{d}t} \right)= 2R{{\dot R}^2} + {R^2}\ddot R.
\end{equation}
{\color{black}The above equation is in exactly the same form as that in the previous work \cite{zhang2023unified}. ${\mathrm{d}F}/{\mathrm{d}t}$ is an operator that can be used to consider the effects of multiple bubbles, boundaries, and the flow field environment. 
}

Let ${\rm {Ma}}=\dot{R}/C$, then equation (\ref{eq-core-2}) can be expressed in the form of Mach number 

\begin{equation}\label{eq-core-3}
\left( { \frac{{1 - {\rm {Ma}}}}{R} + \frac{1}{C}\frac{\mathrm{d}}{{\mathrm{d}t}}} \right)\left( \frac{{R}}{1 - {\rm {Ma}}}\frac{\mathrm{d}F}{\mathrm{d}t} \right)= 2R{{\dot R}^2} + {R^2}\ddot R
\end{equation}

\noindent or

\begin{equation}\label{eq-core-4}
\left( { \frac{{1 - {\rm {Ma}}}}{R} + \frac{\rm Ma}{\dot R}\frac{\mathrm{d}}{{\mathrm{d}t}}} \right)\left( \frac{{R}}{1 - {\rm {Ma}}}\frac{\mathrm{d}F}{\mathrm{d}t} \right)= 2R{{\dot R}^2} + {R^2}\ddot R.
\end{equation}

To obtain the specific expression of ${\mathrm{d}F}/{\mathrm{d}t}$ in equation (\ref{eq-core-2}), the Bernoulli equation is employed:

\begin{equation}
\label{eq-bernoulli}
\frac{ \partial \varphi}{ \partial t} + \frac{1}{2}{\left| \boldsymbol{u} \right|^2} + H = 0,
\end{equation}
\noindent where $H$ is the enthalpy difference; $H=\int_{{P_{\rm{a}}}}^{{P_{\rm{b}}}} {\rho^{-1}} {} \mathrm{d} p$, $P_{\rm b}$ is the fluid pressure on the outside of the bubble surface; ${P_\mathrm{a}}$ is the ambient pressure at the bubble position (${P_\mathrm{a}}={P_\mathrm{B}}+{P_\mathrm{E}}$; ${P_\mathrm{B}}$ is induced by boundaries or other bubbles; ${P_\mathrm{E}}$ denotes extra pressures including the hydrostatic pressure); $\rho$ is the liquid density. In this study, we reserve only the first-order term of $H$ with respect to $(P_\mathrm{b}-P_\mathrm{a})/(\rho C^2)$ and then $H=(P_{\rm b}-P_{\rm a})/\rho$.

{On the bubble surface, the normal stress balance can be expressed as follows:
\begin{equation}\label{eq-normal-stress-balance}
	P_{\rm{g}} = P_{\rm{b}} + \frac{2 \sigma}{R} + \tau_n,
\end{equation}
where $P_{\rm{g}}$, $\sigma$, and $\tau_n$ are the inner bubble pressure, surface tension, and normal viscous stress.
$P_{\rm{g}}$ is calculated with the equation of state of the internal gas-the adiabatic ideal gas equation was adopted in this study for simplicity, as suggested by Lee \cite{Lee2007} and Wang \cite{wb10}. 
According to the Stokes assumption on the viscous stress of Newtonian fluid, the normal viscous stress can be expressed as follows:
\begin{equation}\label{eq-tau-n}
	\tau_n = 2\mu\left(\frac{\partial u_r}{\partial r} -\frac13 \nabla\cdot \boldsymbol{u}\right),
\end{equation}
where $\mu$ is the fluid viscosity. Retaining only the effect of the point source, we have the following relation evaluated at the bubble surface
\begin{equation}
	\frac{\partial u_r}{\partial r} = \frac{\partial ^2 \phi}{\partial r^2} = -2\frac{\dot{R}}{R} + \frac{1}{RC^2}f''.
\end{equation}
As for the velocity divergence term in equation (\ref{eq-tau-n}), employing the linearized equation of the state of the external fluid, we have
\begin{equation}
	\nabla\cdot \boldsymbol{u} = -\frac{1}{\rho C^2}\frac{\partial p}{\partial t}.
\end{equation}
Thus, plug the above two equations into equation (\ref{eq-tau-n}), we have
\begin{equation}
	\tau_n = \frac{4}{R}\mu\dot{R} + O(C^{-2}).
\end{equation}
When the high-order terms are dropped, equation (\ref{eq-normal-stress-balance}) agrees with that used in Zhang et al. \cite{zhang2023unified}.}

Similar to the work of Zhang et al. \cite{zhang2023unified}, ${\mathrm{d}F}/{\mathrm{d}t} $ can be derived as follows:
\begin{equation}
\label{cm20}
\frac{\mathrm{d}F}{\mathrm{d}t} 
=  R \left( {1 - \frac{{\dot R}}{C}} \right) \left( {\frac{1}{2}{{\dot R}^2} + \frac{1}{4}v^2 + H} \right).
\end{equation}

Substituting equation (\ref{cm20}) into equation (\ref{eq-core-2}), the bubble oscillation equation is obtained: 
\begin{equation}
\label{eq33}
\left( {\frac{{C - \dot R}}{R} + \frac{\mathrm{d}}{{\mathrm{d}t}}} \right)\left[ {\frac{{{R^2}}}{C}\left( {\frac{1}{2}{{\dot R}^2} + \frac{1}{4}{v^2} + H} \right)} \right]  = 2R{{\dot R}^2} + {R^2}\ddot R.
\end{equation}

Expanding the above equation, we could get the following differential equation:
\begin{multline}
\left(1+\frac{ \dot{R}}{C} \right) H+
\frac{R}{ C} \dot{H}
+\frac14 \left(1+\frac{\dot{R}}{C}\right)v^2 + \frac{{R}}{2C}v\dot{v} \\
=
\frac{3}{2} \left(1-\frac{ \dot{R}}{3C} \right)\dot{R}^2 +
\left(1-\frac{ \dot{R}}{C} \right) R\ddot{R},
\label{Equation:Keller1}
\end{multline}
which includes the effect of the compressible bubble migration on bubble pulsation that is ignored in the Keller-Miksis equation. {\color{black} Equation (\ref{Equation:Keller1}) is a typical second-order ordinary differential equation, which is often solved using the fourth-order Runge-Kutta method to obtain the temporal variation of the bubble radius.}

Another variable $v$ in equation (\ref{eq33}) is obtained using the bubble migration equation \cite{zhang2023unified}:
\begin{equation}
\label{eq-migration}
{C_\mathrm{a}} R \dot {\boldsymbol{v}} + \left( {3{C_\mathrm{a}}\dot R + {{\dot C}_\mathrm{a}}R} \right) \boldsymbol{v} 
+ \frac{R}{\rho }\nabla {P_\mathrm{a}} + \frac{3}{8}{C_\mathrm{d}}\mathbb{S} \left(\boldsymbol{v}\right) = 0,
\end{equation}
\noindent where $C_\mathrm{a}$ and $C_\mathrm{d}$ are the added mass coefficient and drag coefficient, respectively; the values of $C_\mathrm{a}$ and $C_\mathrm{d}$ are taken as 1.0 and 0.5, respectively; $\mathbb{S}(\cdot)=(\cdot)|\cdot|$. 

To include the propagation of pressure waves \cite{zhang2023unified}, the pressure of a point $\bm r$ in the flow field induced by a bubble is calculated as follows:
\begin{align}
\label{eq25}
p(\boldsymbol{r},t) &=  P_\mathrm{a} +  \rho \left.
\left\{
\frac{R}{|\boldsymbol{r}|}\left(H+\frac12\dot{R}^2\right) 
\notag\right. \notag\right.
\\&
\phantom{=\;\;}
\left.  \phantom{=\;\;}
\left. 
-\frac12\frac{R^2}{|\boldsymbol{r}|^4}
\left[R\dot{R}+\frac{|\boldsymbol{r}|-R}{C}\left(H+\frac12 \dot{R}^2\right)
\right]^2
\right\}
\right|_{\left(R,t-\frac{|\boldsymbol{r}|-R}{C}\right)}. 
\end{align}

\section{Modeling bubble dynamics near hybrid boundaries}\label{sect:theory-crossboundary}

\subsection{Bubble dynamics at a corner crossed by two boundaries}
A type of common hybrid boundary is the corner. Near a corner formed by two crossed boundaries, bubble dynamics are influenced by the boundaries in two distinct directions. Predicting bubble behavior near corners at arbitrary angles is challenging due to the lack of physical mapping between corners at different angles. However, {\color{black} when dealing with two semi-infinite boundaries at an angle of $\alpha=\pi/k$ (where $k$ is a positive integer), the boundary effect can be accurately modeled by incorporating a finite number of mirror bubbles with a specific spatial distribution based on image theory.}

\begin{figure*}
	\centering\includegraphics[width=0.8\linewidth]{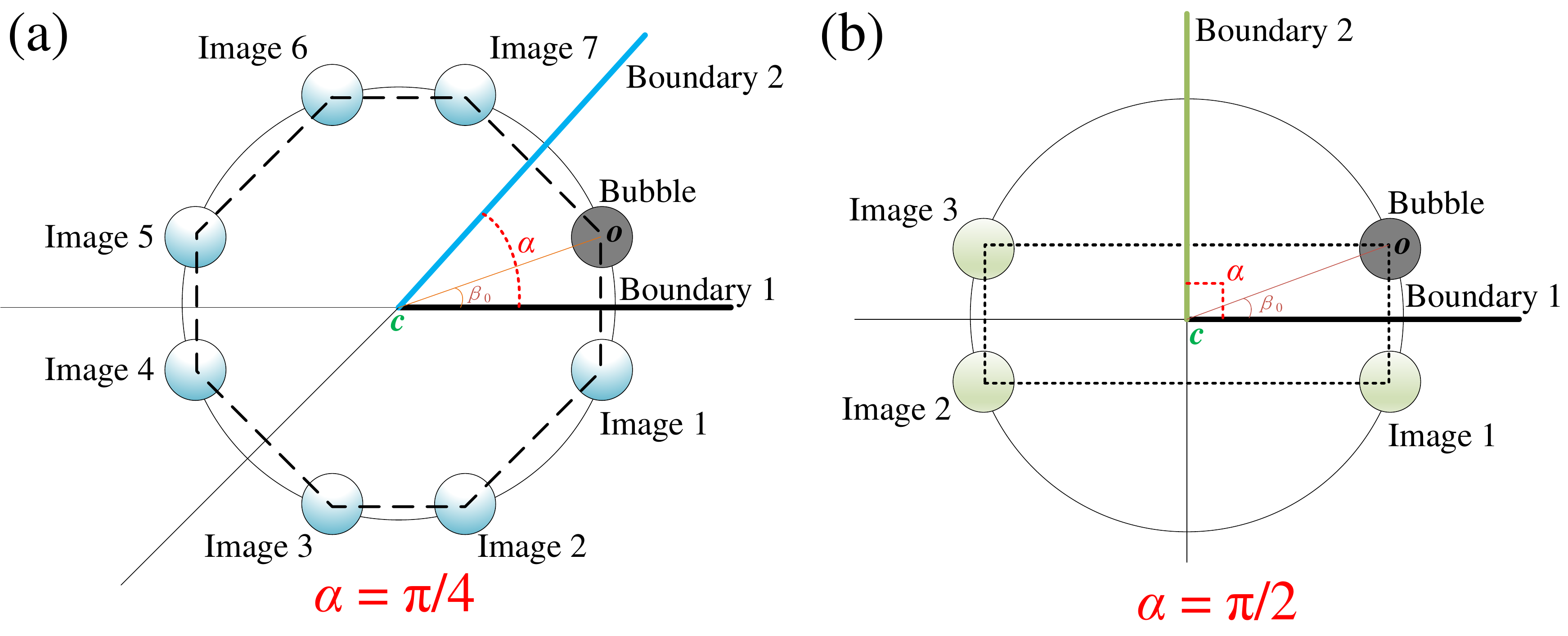}
	\caption{Schematic diagram depicting the image theory and mirror bubbles for the crossed boundaries with two angles  (a) $\alpha=\pi/4$ (b) $\alpha=\pi/2$. The black circle denotes the real bubble, and the others denote mirror bubbles. The solid lines represent the boundaries. The origin of the coordinate system is at the initial bubble center and the direction of bubble migration is consistent with $\theta=0$, which is the same as that in Fig. 1. }
	\label{crossed}
\end{figure*}

To account for the effects of two boundaries using image theory, the positions of mirror bubbles on both sides of any boundary need to be completely symmetrical, which requires the position of the mirror bubbles to be arranged reasonably. The spatial distributions of mirror bubbles for two typical corners of two walls ($\alpha= \pi/4$ and $\pi/2$) are presented in Fig. \ref{crossed}. The coordinate system is the same as that in Fig. 1, in which the origin is set at the initial bubble center and $\theta=0$ denotes the direction of bubble migration. The black circle represents the real bubble while others denote mirror bubbles. Define $\bm o_N$ as the position vector of the real bubble $N$. Let us describe the plane of boundary $i$ with ${{{\bm n}_i}} \cdot {\bm r} +b_i=0$, where ${\bm n}_i$ denotes the outer normal vectors of boundary $i$ and $b_i$ is a constant. The position of each mirror bubble is obtained by the symmetry of the other bubbles about the boundary. For example, mirror bubbles 1 and 2 are obtained from the real bubble symmetric about boundaries 1 and 2, respectively; mirror bubbles 3 and 4 are obtained from mirror bubbles 1 and 2 symmetric about boundaries 2 and 1, respectively. Here, we denote the position vector of the mirror bubble $i$ of bubble $N$ as ${{\bm o}}_{Ni}$: 
\begin{small}
	\begin{equation}
	\label{jiao1}
	{{\bm o}_{Ni}} = \left\{ {\begin{array}{*{20}{c}}
		{{{\bm o}_{N({i - 1})}} - \left( {2{{\bm o}_{N({i - 1})}} \cdot {{\bm n}_1}} +b_1 \right){{\bm n}_1}{\rm{   \quad \quad  when\ }} i\ {\rm{ is\  odd}}}\\
		{{{\bm o}_{N[{\max \left( {i - 3,0} \right)}]}} - \left( {2{{\bm o}_{N[{\max \left( {i - 3,0} \right)}]}} \cdot {{\bm n}_2}} +b_2\right){{\bm n}_2}{\rm{ \quad when\ }} i{\rm{\ is\ even}}}
		\end{array}} \right.,
	\end{equation}
\end{small}
\noindent in which $i = 1,2,...,2k-1$; ${{\bm o}_{N0}} = {\bm o_N}$.  

Let the reflection coefficients of boundaries 1 and 2 be $\omega_1$ and $\omega_2$, respectively. The reflection coefficient is 1.0 for the rigid boundary and -1.0 for the ideal free surface. Thus,  the coefficient of the mirror bubble $\xi _{Ni}$ can be expressed as follows:
\begin{equation}
\label{jiao2}
{\xi _{Ni}} = \left\{ {\begin{array}{*{20}{c}}
	{{\xi _{N(i - 1)}}{\omega _1}{\rm{\quad \quad  \quad       when\ }}i{\rm{\ is\ odd}}}\\
	{{\xi _{N[\max \left( {i - 3,0} \right)]}}{\omega _2}{\rm{ \quad  when\ }}i{\rm{\ is\ even}}}
	\end{array}} \right..
\end{equation}
in which ${\xi _{N0}} = 1$.

As the bubble dynamics are affected by all the image bubbles, equations of multi-bubble interaction are employed \cite{zhang2023unified} to consider the influence of crossed boundaries. The velocity $\boldsymbol{u}$ and derivative of velocity potential $\dot{\varphi}$ induced by the bubble $N$ at $\bm r$ are as follows:
\begin{small}
\begin{equation}
\label{sc}
\boldsymbol{u}_{(\boldsymbol{r},t)} = 
-\frac{\boldsymbol{o}_N - \boldsymbol{r}}{|\boldsymbol{o}_N - \boldsymbol{r}|^3}\left.
\left[\frac{R_N}{C}\left(|\boldsymbol{o}_N - \boldsymbol{r}|-R_N\right)\left(H_N+\frac12\dot{R}_N^2\right)	
+R_N^2\dot{R}_N
\right]
\right|_{\left({R_N},t_c\right)}
\end{equation}
\end{small}

\noindent and

\begin{equation}
\label{pc}
\dot{\varphi}_{(\boldsymbol{r},t)} = 
-\left. \frac{R_N}{|\boldsymbol{o}_N - \boldsymbol{r}|} \left( H_N + \frac12 \dot{R}_N^2\right) \right|_{\left({R_N},t_c\right)},
\end{equation}
\noindent in which $t_c=t-(|{\bm o}_N-{\bm r}|-R_N)/C$.

Considering the effect of mirror bubbles and other real bubbles, the velocity $\boldsymbol{u}_{\mathrm{B}}$ and pressure $P_{\mathrm{B}}$ of the background flow field of real bubble $M$ can be written as follows:
\begin{multline}
\label{eq47}
\boldsymbol{u}_{\mathrm{B}}{\left( {{{\boldsymbol{o_M}}},t} \right)}= \sum_{ N = 1,L\atop N \ne M} \boldsymbol{u}_N(\boldsymbol{o_M},t) \\+  \sum_{N=1,L} \sum_{i=1,2k-1} \xi _{Ni} \boldsymbol{u}_{{Ni}}{(\boldsymbol{o_M},t)}
\end{multline}

\noindent and

\begin{equation}
\begin{aligned}
\label{eq-pa-boundary}
P_{\mathrm{B}}{(\boldsymbol{o_M},t)} =&-\rho \sum_{N = 1,L\atop N \ne M} \dot{\varphi}_{{N}}{(\boldsymbol{o_M},t)}
\\&- \rho \sum_{ N = 1,L} \sum_{i=1,2k-1} \xi_{Ni} \dot{\varphi}_{{Ni}}{(\boldsymbol{o_M},t)}
-\frac12 \rho |\boldsymbol{u}_{\mathrm{B}}{(\boldsymbol{o_M},t)}|^2,
\end{aligned}
\end{equation}
respectively. $L$ is the number of the real bubbles. 

By substituting the above two equations
into equations (\ref{eq33}) and (\ref{eq-migration}) for the real bubble, we have the equations for the bubble dynamics considering the crossed boundary effect in the compressible flow. 

Then, the pressure in the flow field is the sum of that induced by the real bubble and all the mirror bubbles:
\begin{footnotesize}
	\begin{align}
	\label{pretotal}
	p(\boldsymbol{r},t) &=  P_\mathrm{a} +  \rho \sum_{N=1,N_{all}} \left.
	\left\{
	\frac{R_N}{|{\bm o}_N-\boldsymbol{r}|}\left(H_N+\frac12\dot{R_N}^2\right) 
	\notag\right. \notag\right.
	\\&
	\phantom{=\;\;}
	\left.  \phantom{=\;\;}
	\left. 
	-\frac12\frac{R_N^2}{|{\bm o}_N-\boldsymbol{r}|^4}
	\left[R_N\dot{R_N}+\frac{|\boldsymbol{r}_N|-R_N}{C}\left(H_N+\frac12 \dot{R_N}^2\right)
	\right]^2
	\right\}
	\right|_{\left(R_N,tc\right)},
	\end{align} 
\end{footnotesize}
in which $N_{all}$ is the number of all bubbles including real and mirror bubbles.

As $C$ tends to infinity, the above model reduces to the bubble dynamics equations for bubble $M$ in the incompressible fluids:

\begin{multline}
\label{eq43}
\frac{1}{4}{\left| {{{\boldsymbol{v}}_M} + \sum_{ N = 1,2k-1\atop
			N \ne M} {{{\boldsymbol{r}}_{MN}}\frac{{R_N^2{{\dot R}_N}}}{{{{\left| {{{\boldsymbol{r}}_{MN}}} \right|}^3}}}} } \right|^2}  
- 
\sum_{ N = 1,2k-1\atop
	N \ne M} {\left( {\frac{{2{R_N}\dot R_N^2 + R_N^2{{\ddot R}_N}}}{{{{\left |{\boldsymbol{r}}_{MN} \right | }}}}} \right)} \\
+\frac{{{P_{\rm{b}}} - {P_\mathrm{E} }}}{\rho } = 
\frac{3}{2}{\dot R_M^2} + R_M\ddot R_M,
\end{multline}
\noindent and
\begin{multline}\label{eq44}
\sum_{ N = 1,2k-1\atop
	N \ne M} {{{\boldsymbol{r}}_{MN}}\frac{{(2+5C_{\rm a})R_N^2\dot R_N^2 + (1+C_{\rm a})R_N^3{{\ddot R}_N}+\dot{C_{\rm{a}}} R_N^3 {\dot R_N} }}{{{{\left| {{{\boldsymbol{r}}_{MN}}} \right|}^3}}}} \\-
\frac38{C_{\rm{d}}}\mathbb{S} 
\left( {{{\boldsymbol{v}}_{M}} + \sum_{ N = 1,2k-1\atop
		N \ne M} {{{\boldsymbol{r}}_{MN}}\frac{{R_N^2{{\dot R}_N}}}{{{{\left| {{{\boldsymbol{r}}_{MN}}} \right|}^3}}}} } \right) 
+ R{\boldsymbol{g}}= C_{\rm{a}}R {\dot{\boldsymbol{ v}}_M} + (3C_{\rm{a}}\dot R + \dot{C_{\rm{a}}}R){{\boldsymbol{v}}_M}.
\end{multline}

Equation (\ref{eq44}) reduces to the incompressible multi-bubble equation, as presented in the previous work \cite{zhang2023unified} when $C_{\rm a}$ = 0.5  on the left side.

\subsection{Bubble dynamics between two parallel boundaries}\label{sect:theory-parallelboundary}

\begin{figure*}
	\centering\includegraphics[width=0.6\linewidth]{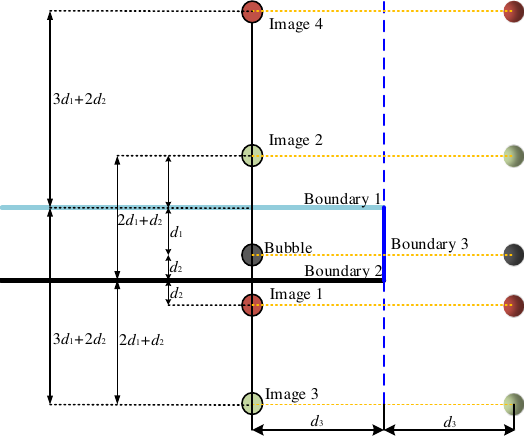}
	\caption{Schematic diagram depicting the mirror bubbles set to simulate the boundary effect containing crossed and parallel boundaries. Three solid lines represent three boundaries with boundary 3 perpendicular to the other two boundaries. The black circles on the left side of boundary 3 denote the real bubble, while the rest are mirror bubbles distributed alternately about boundaries 1 and 2. Circles on the right side of boundary 3 denote the mirror bubbles of all bubbles on the left side.}
	\label{parallel}
\end{figure*}

As the angle of the crossed boundary is zero, the two boundaries become parallel and ${\bm n}_1+{\bm n}_2=0$, as shown by boundaries 1 and 2 in Fig. \ref{parallel}. Using image theory to simulate the effect of two parallel boundaries on a bubble in-between presents a challenge, as an infinite number of mirror bubbles is theoretically required to represent the boundary condition (the maximum value of $i$ in equations (\ref{jiao1}) and (\ref{jiao2}) tends to infinity). To overcome this challenge, we gradually increased the number of mirror bubbles and calculated the dynamic behavior of the real bubble. We continued this process until the calculation results reached a steady state, where further increases in the number of mirror bubbles no longer significantly affected the bubble dynamics results. The number of mirror bubbles is defined as $m$. At this point, we deemed the reproduction of the two parallel boundaries' effect to be satisfactory. We define the upper boundary as boundary 1 and the lower boundary as boundary 2.

\begin{figure}[H]
	\centering\includegraphics[width=1.0\linewidth]{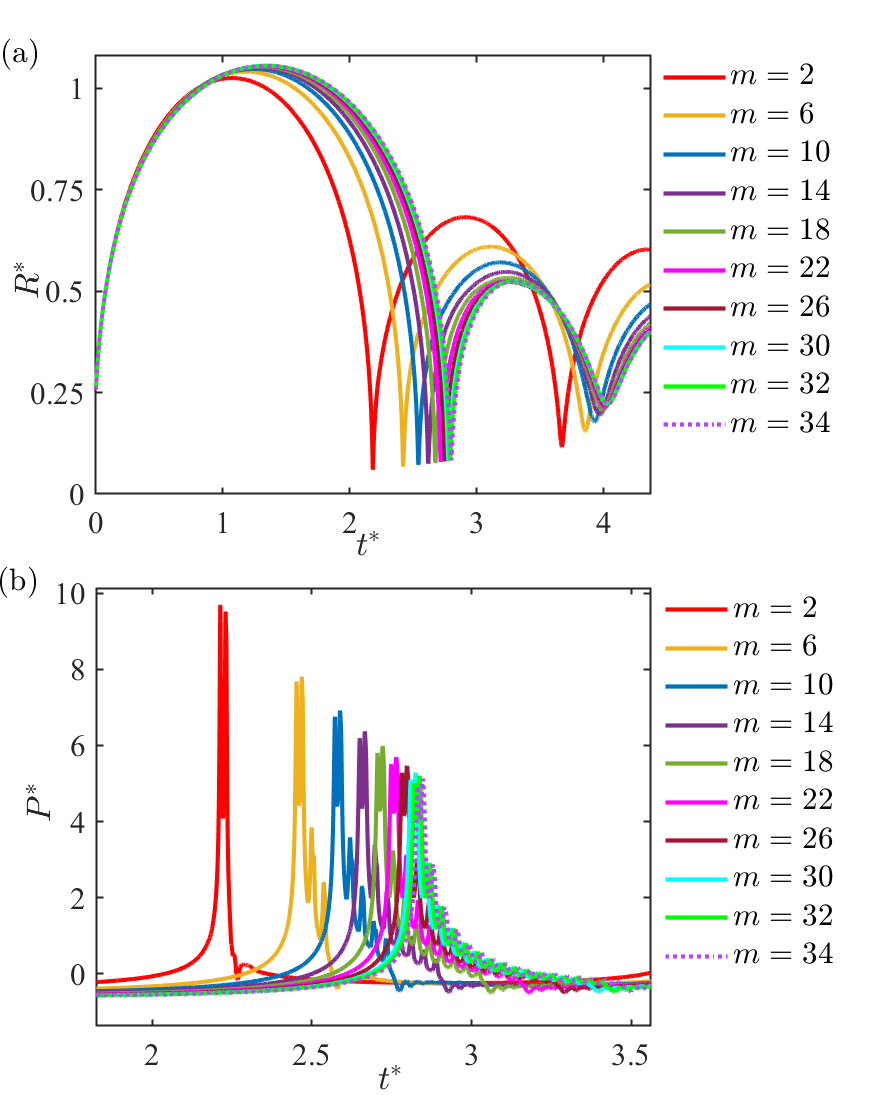}
	\caption{Convergence of (a) the bubble radius and (b) the flow field pressure with the number of mirror bubbles. The bubble is between two infinite parallel rigid walls (boundaries 1 and 2). The dimensionless distance between the two boundaries is 5.0 with the initial bubble 3.0 away from boundary 1, and the dimensionless initial conditions for the bubble are $P_0^* = 20$, $ R_0^* = 0.06 $ and $ \dot{R}_0^*= 5.98$.}
	\label{convegency}
\end{figure}

To illustrate the convergence of the result as the number of mirror bubbles is sufficiently large, a typical case is considered with $\omega_1$ and $\omega_2$ set to 1.0. The results are non-dimensionalized with the maximum radius of the bubble in the free field, fluid density ($\rho=1000$ kg/$\rm m^3$), and pressure at infinity as reference values. The initial conditions for bubble pulsation are $P_0^* = 20$, $ R_0^* = 0.06$ and $ \dot{R}_0^* = 5.98$, in which the superscript '*' represents the dimensionless physical quantity. The initial dimensionless distance between the bubble and boundary 1 is 3.0, while the distance from boundary 2 is 2.0. The number of mirror bubbles $m$ ranges from 1 to the case where the peak pressure of bubble pulsation converges, as shown in Fig. \ref{convegency}. As $m$ increases, the asymmetry effect of the mirror bubbles on both sides of the boundaries decreases, and the difference between the calculated results reduces. As $m$ is small, the error of the calculation results is relatively large since the excess mirror bubbles significantly impact the real bubble. If we focus only on the radius and period of the bubble, the calculation results converge within 0.1\% after $m$ reaches 14. After $m$ reaches 30, the peak pressure achieves convergence within a tolerance of 0.1\%. Thus, we take $m$ as 30 in the subsequent calculations. 

\subsection{Bubble dynamics between two parallel boundaries and near a sidewall}

Further, when the boundary contains both crossed and parallel features, more mirror bubbles are involved and the pressure and velocity of the background flow field of bubbles are a linear superposition of mirror bubbles and other real bubbles. For example, when boundary 3 is introduced as illustrated in \ref{parallel}, the positions of the bubbles at both sides of the three boundaries should be perfectly symmetrical. However, when the angle between two of the boundaries is obtuse, the image method cannot equate the boundary effect. Thus, only the situation where the sidewall is perpendicular to the free surface and bottom could be modeled, that is, a semi-truncated rectangular channel. Under this condition we need to symmetrically move all the bubbles under the influence of the two parallel boundaries to the other side of the third boundary, as shown in Fig. \ref{parallel}. Thus, the position vector and coefficient of the mirror bubble on the other side of boundary 3 are as follows:

\begin{equation}
\label{jiao5}
\begin{array}{l}
{{\bm o}_{N(m+i)}} = 2{\bm o}_{Ni} -(2{\bm o}_{Ni} \cdot {\bm n}_3 +b_3 ){\bm n}_3
\end{array}
\end{equation}
and
\begin{equation}
\label{jiao21}
{\xi _{N(m+i)}} = \xi_{Ni} \omega_3,
\end{equation}
\noindent in which ${\bm n}_3$ is the outward normal vector and $\omega_3$ is its reflection coefficient; $b_3$ is a constant; $i$ ranges from 1 to $m$ in equations \ref{jiao5} and \ref{jiao21}. As the bubble $2m+1$ is symmetry with the real bubble about boundary 3, we have ${{\bm o}_{N(2m+1)}}={\bm o}_{N}-(2{\bm o}_N \cdot {\bm n}_3+b_3 ){\bm n}_3$, and ${\xi _{N(2m+1)}} = \omega_3$.

The velocity and pressure of the background flow field of the real bubble $M$ can be written as follows:
\begin{small}
	\begin{multline}
	\label{eq471}
	\boldsymbol{u}_{\mathrm{B}}{\left( {{{\boldsymbol{o_M}}},t} \right)}= \sum_{ N = 1,L\atop N \ne M} \boldsymbol{u}_N(\boldsymbol{o_M},t) +  \sum_{N=1,L} \sum_{i=1,2m+1} \xi _{Ni} \boldsymbol{u}_{{Ni}}{(\boldsymbol{o_M},t)}
	\end{multline}
\end{small}
\noindent and

\begin{small}
	\begin{equation}
	\begin{aligned}
	\label{eq-pa-boundary11}
	P_{\mathrm{B}}{(\boldsymbol{o_M},t)} =&-\rho \sum_{N = 1,L\atop N \ne M} \dot{\varphi}_{{N}}{(\boldsymbol{o_M},t)}
	- \rho \sum_{ N = 1,L} \sum_{i=1,2m+1} \xi_{Ni} \dot{\varphi}_{{Ni}}{(\boldsymbol{o_M},t)}\\&
	-\frac12 \rho |\boldsymbol{u}_{\mathrm{B}}{(\boldsymbol{o_M},t)}|^2.
	\end{aligned}
	\end{equation}
\end{small}

\section{Results and discussion}\label{sect:comparison}
\subsection{Bubble dynamics near hybrid boundaries}
\begin{figure*}
	\centering\includegraphics[width=1.0\linewidth]{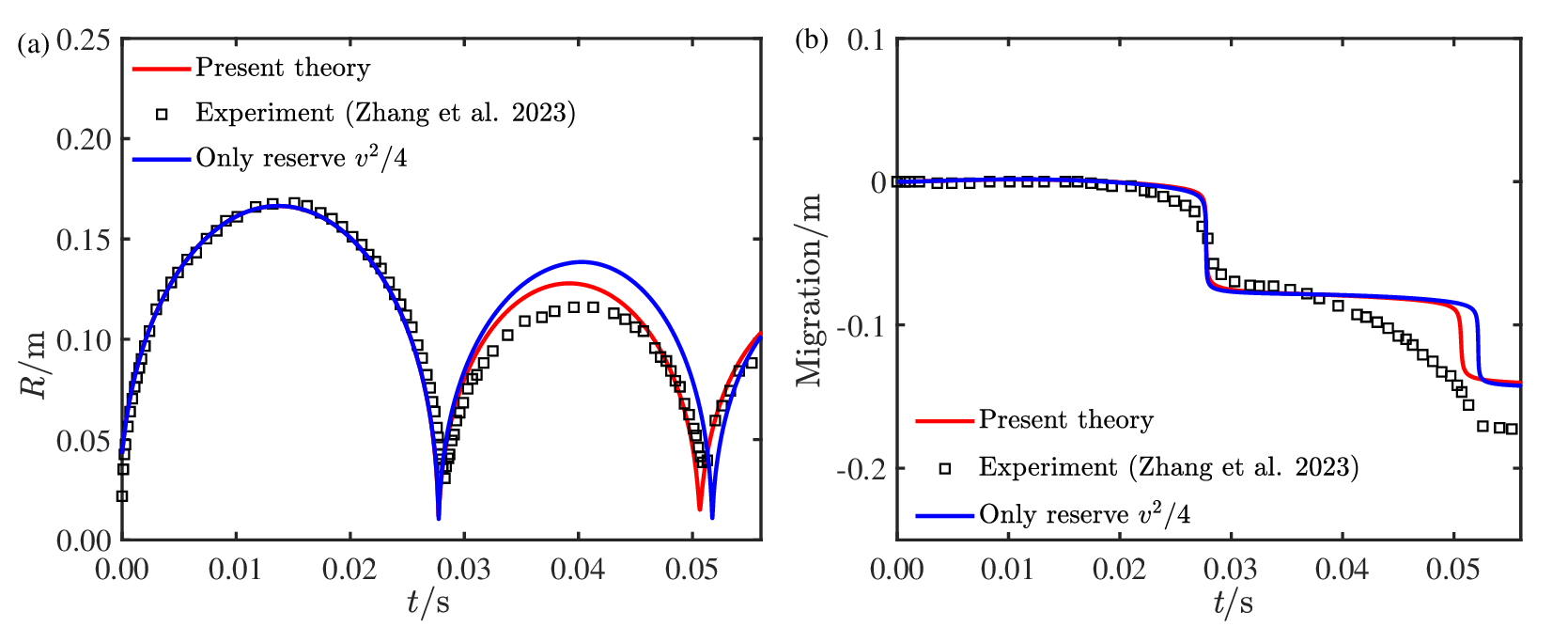}
	\caption{Comparison between the theoretical and experimental results of (a) bubble radius and (b) vertical bubble migration. The bubble is generated by an underwater explosion of 1.125 g PBX at a distance of 0.4 m from the free surface. The experimental values come from the work of Zhang et al. \cite{zhang2023unified}. }
	\label{UEXf}
\end{figure*}

\begin{figure*}
	\centering\includegraphics[width=0.95\linewidth]{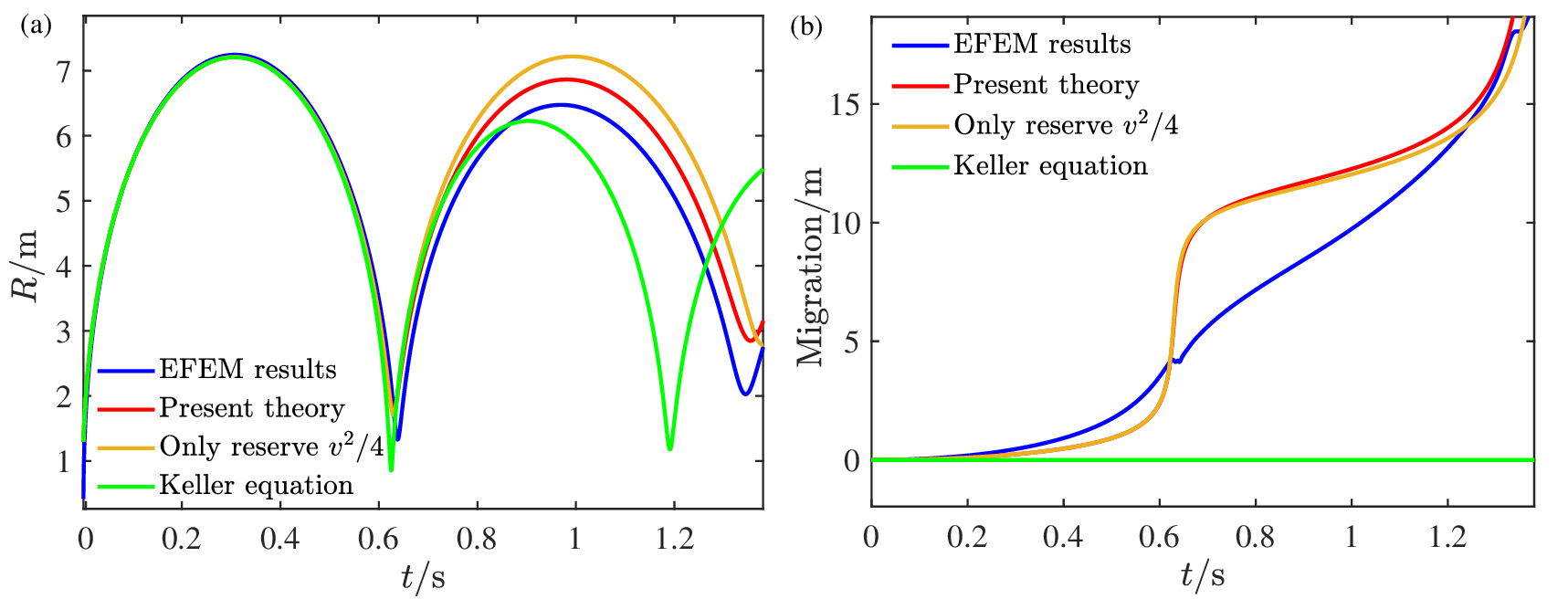}
	\caption{Comparison between the theoretical and numerical results of (a) bubble radius and (b) vertical bubble migration. The bubble is generated by an underwater explosion of 500 Kg TNT explosive at a depth of 40 m. The effect of the free surface is ignored in both theoretical and numerical results as the distance between the bubble and free surface exceeds 5 times the maximum bubble radius.}
	\label{CELcom}
\end{figure*}

In the work of Chahine et al. \cite{Chahine2009,Ma18J}, the Keller equation was modified by adding $v^{2}/4 $ to account for the effect of bubble migration on bubble pulsation, similar to the bubble dynamics equation in the incompressible fluid \cite{h70,Best1991}. However, this approach does not consider the effect of compressible bubble migration.  In this study, we investigated the effect of bubble migration on bubble oscillation in a compressible flow by removing the terms ${{\dot R}}v^2/{4C}$ and ${{R}}v\dot{v}/{2C}$ in equation (\ref{Equation:Keller1}) and compared the results with the original equation. Figure \ref{UEXf} presents a comparison between theory and experiment, where a bubble was generated by the explosion of 1.125 g PBX at a distance of 0.4 m from the free surface. Further details on the experimental system and the initial calculating conditions can be found in the previous work \cite{zhang2023unified}. The red solid line shows the result of the original equation, while the blue line represents the result when only the term $v^2/4$ is retained among the three migration-related terms. The radii and period of the red and blue lines differ significantly, especially in the second cycle of the bubble, indicating that the compressible terms ${{\dot R}}v^2/{4C}$ and ${{R}}v\dot{v}/{2C}$ affect the energy loss of the bubble caused by the fluid compressibility. The deviation between the theoretical migrations and the experiment occurs mainly in the second and later cycles and could be due to complex factors, including the non-spherical bubble behavior induced by gravity or boundaries, such as jetting, splitting, and phase transition.
In addition, we also calculated the time histories of the radius and migration for a large-charge underwater explosion bubble, and compared them with the results from the Euler finite element method (EFEM) \cite{Tian2020OE,LiuNN2020}, as shown in Fig. \ref{CELcom}. EFEM is a numerical method used to solve the compressible Euler equation system. It proves to be effective in simulating deformable bubbles. The bubble is generated by 500 Kg TNT explosives at a depth of 40 m. Compared to the result obtained by directly adding $v^2/4$ and the result from Keller equation, the bubble radius and migration in the present theory agree better with the numerical simulation results. The two terms related to compressible migration obviously affect the dynamic characteristics of the bubbles. The result in the present theory agrees well with that of EFEM, especially in the second cycle. Therefore, the present theory, with all three migration-related terms, provides a more accurate approach for accounting for the effect of bubble migration on bubble pulsation in compressible fluids simultaneously.

\begin{figure*}
	\centering\includegraphics[width=1.0\linewidth]{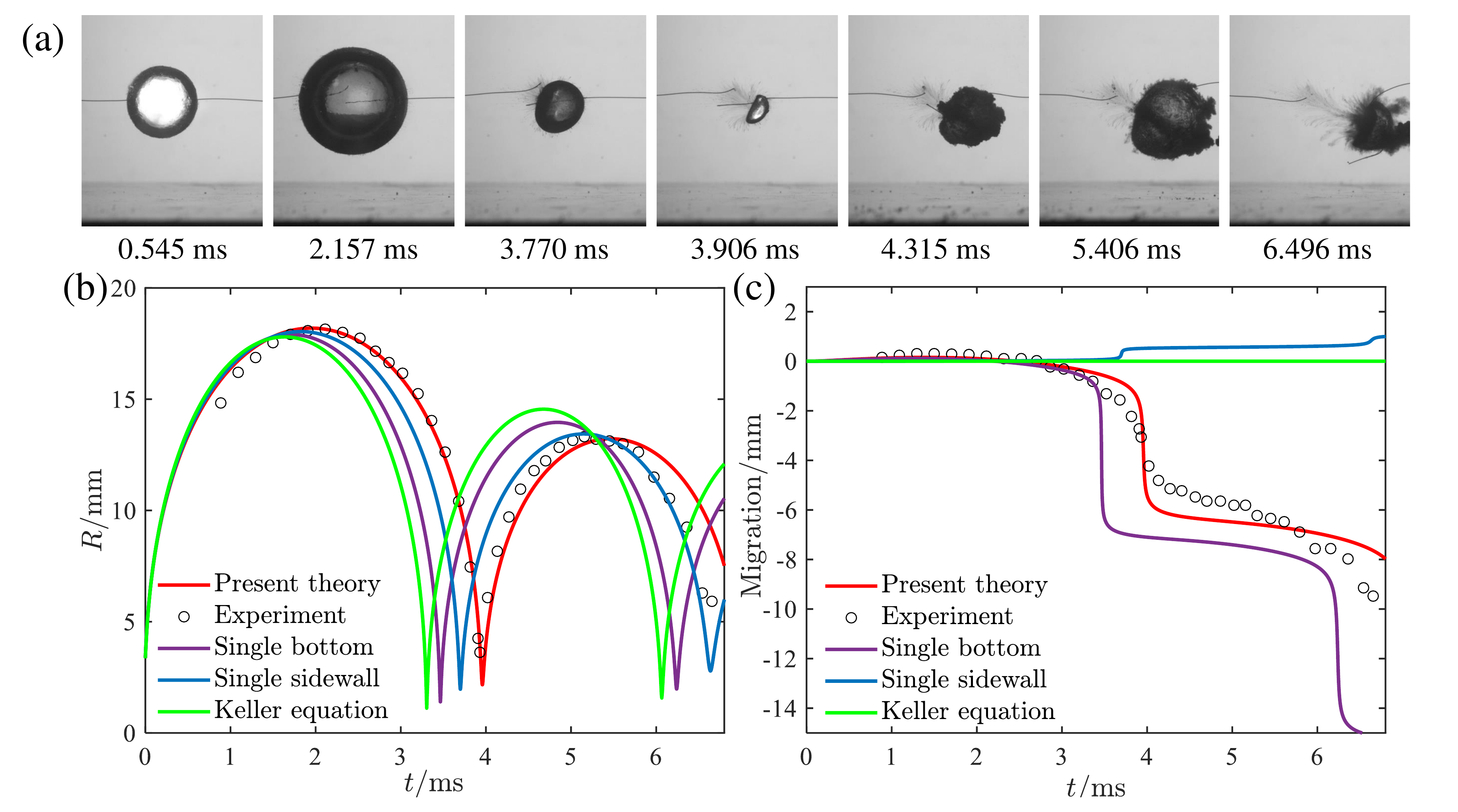}
	\caption{Spark-induced bubble experiment and its comparisons of theoretical and experimental results near the bottom and vertical sidewall. The vertical sidewall on the right side is equivalently represented by two horizontally arranged spark-induced bubbles of the same size in the experiment. The distances of the initial bubble from the bottom and sidewall are 34 mm and 33 mm, respectively. (a) Experimental images at different moments. Frame width: 64 mm. (b) Bubble radius (in theoretical calculations: $P_0 = 1.2$ MPa, $ R_0 = 3.35 $ mm and $ \dot{R}_0= 95$ m/s.). (c) Vertical bubble migration.}
	\label{bbR}
\end{figure*}

\begin{figure*}
	\centering\includegraphics[width=1.0\linewidth]{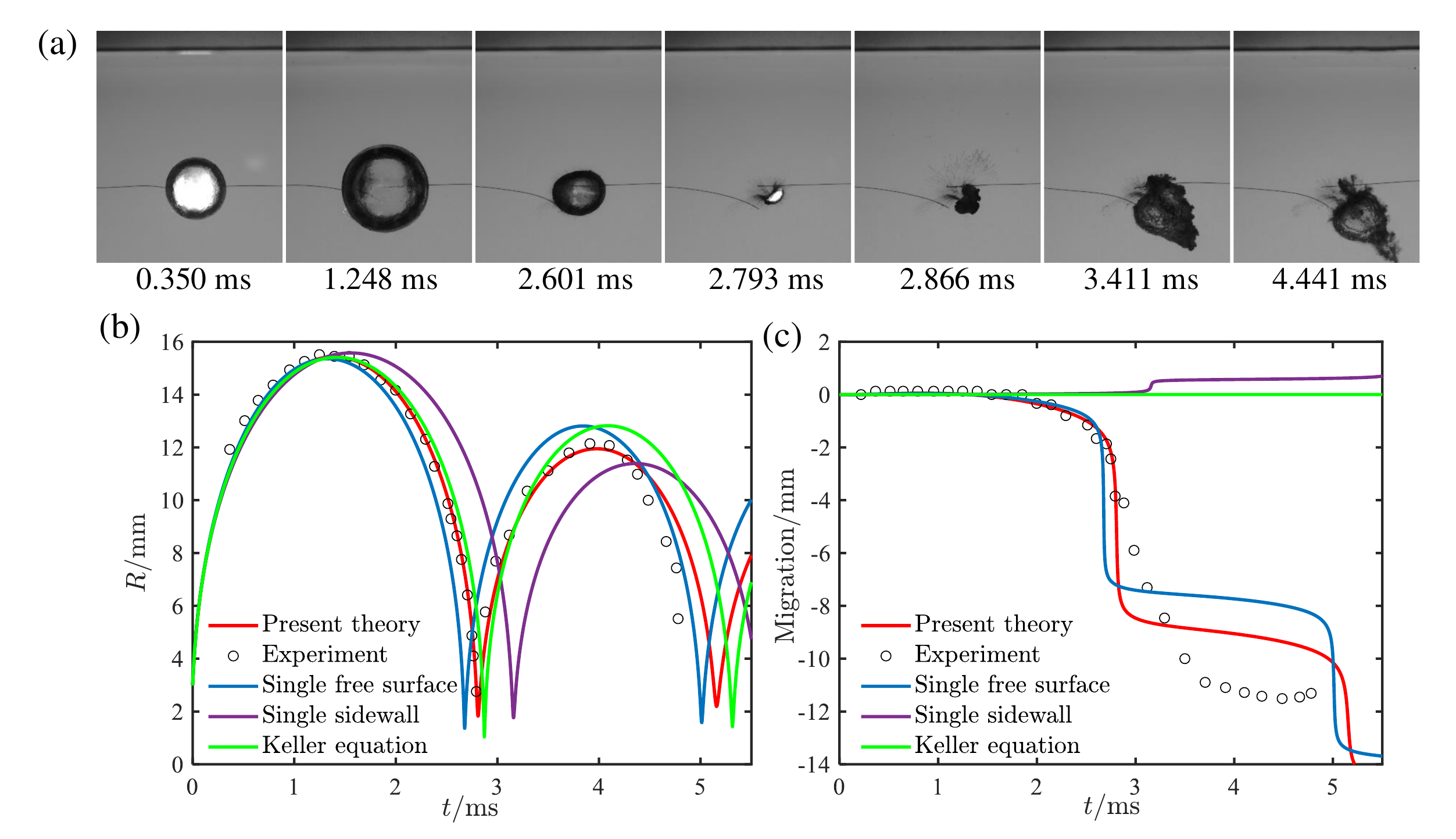}
	\caption{Spark-induced bubble experiment and comparisons of theoretical and experimental results near the free surface and vertical sidewall. Two bubbles of the same size are generated in the experiment to equate a vertical sidewall in the mid-pendant of the two bubbles. The distances of the initial bubble from the free surface and sidewall are 49.5 mm and 35 mm, respectively. (a) Experimental images at different moments. Frame width: 65 mm. (b) Bubble radius (in theoretical calculations: $P_0 = 1.2$ MPa, $ R_0 = 2.98 $ mm and $ \dot{R}_0= 90$ m/s.). (c) Vertical bubble migration.}
	\label{fbR}
\end{figure*}

To validate the present theory in predicting bubble dynamics near hybrid boundaries, we conducted two experiments involving spark-induced bubbles and compared the bubble radii and vertical migration between theory and experiment, as illustrated in Fig. \ref{bbR} and Fig. \ref{fbR}. In the first case, we calculated the bubble dynamics near two perpendicular walls. In the experiment, to account for the effect of the vertical sidewall, we simultaneously generated two bubbles of the same size to equate a vertical rigid wall in the mid-pendant of the two bubbles, according to image theory. The initial bubble is located 33 mm from the sidewall surface and 34 mm from the bottom. 
{\color{black}The initial conditions for spark-induced bubbles in the theory are determined using the method proposed by Wang \cite{w13}: the maximum radius $R_{\rm{max}}$ and the minimum pressure $P_{\rm{min}}$ are required at the moment of maximum bubble volume. $P_{\rm min}$ can be determined by the ratio $R_{\rm max2}/R_{\rm max}$ where $R_{\rm max2}$ is the maximum bubble radius in the second cycle. We performed a backward integration starting from the moment of maximum bubble volume with zero wall velocity until the initial condition is obtained.}
The initial conditions of the bubble in this case are as follows: $P_0 = 1.2$ MPa, $ R_0 = 3.35 $ mm and $ \dot{R}_0= 95$ m/s. We also provided time-history curves of the bubble radius and migration when the sidewall and bottom are present individually. The combined effect of the two walls significantly prolongs the bubble pulsation period, leading to a significant deviation of Keller equation in predicting the bubble period. Our theory is in excellent agreement with the experimental results.

In the second case, we compute the bubble dynamics near the free surface and vertical sidewall at a distance of 49.5 mm from the free surface and 35 mm from the sidewall. Similarly, the sidewall is equivalently represented by two horizontally arranged spark-induced bubbles of the same size in the experiment. The initial conditions of the bubble are $P_0 = 1.2$ MPa, $ R_0 = 2.98 $ mm and $ \dot{R}_0= 90$ m/s. Considering only the effect of the free surface underestimates the pulsation period of the bubble, while considering only the sidewall overestimates it. The vertical migration of the bubble is mainly caused by the free surface. Due to the neglect of boundary effects and bubble migration, Keller equation cannot accurately calculate the period and energy loss of the bubble. Our theoretical results reproduce the evolution of the bubble radius and migration in the experiment well.

\begin{figure*}
	\centering\includegraphics[width=1.0\linewidth]{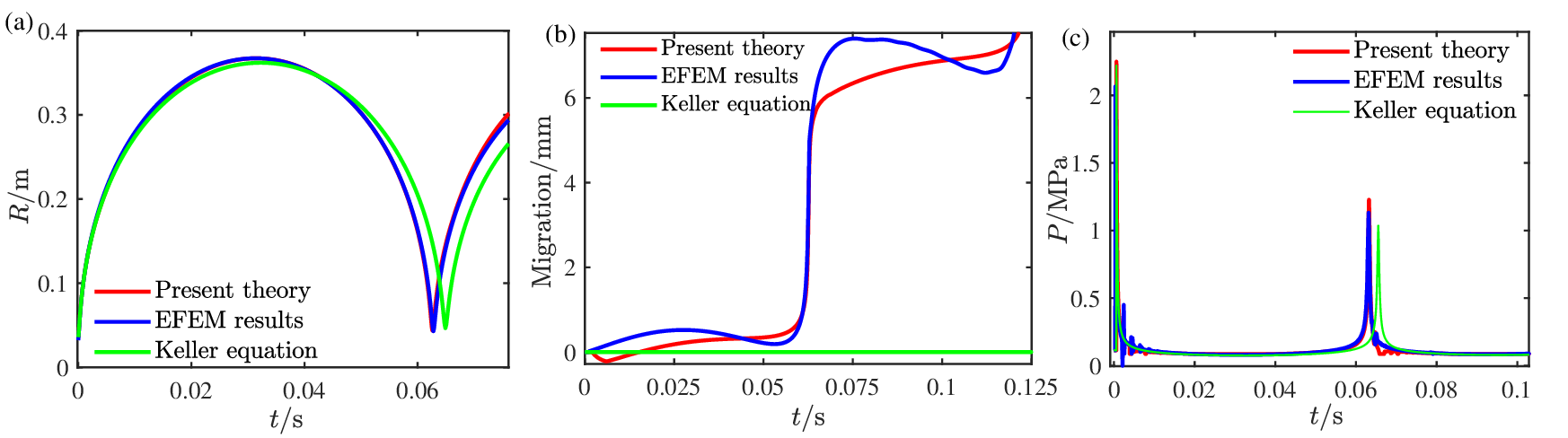}
	\caption{Comparison between the theoretical and EFEM simulation results for a bubble between the free surface and bottom. The distance between the two boundaries is 3 m and the initial bubble is equally distant from the free surface and bottom. In both theoretical and numerical calculations: $P_0 = 60$ MPa, $ R_0 = 32 $ mm and $ \dot{R}_0= 0$ m/s. (a) Bubble radius. (b) Vertical bubble migration. (c) Pulsation pressure (the measurement point is set at 1 m away from the initial bubble in the horizontal direction).}
	\label{parallelbfCEL}
\end{figure*}

To validate the influence of parallel boundaries in our theory, we compared them with numerical simulations using EFEM. The results are shown in Figure \ref{parallelbfCEL}, which includes the bubble radius, migration, and pulsation pressure. The bubble is initially located between two parallel boundaries, namely the free surface and the bottom wall, with a distance of 1.5 m from both boundaries. The initial pulsation conditions of the bubble are: $P_0 = 60$ MPa, $ R_0 = 32 $ mm, and $ \dot{R}_0= 0$ m/s. The maximum radius of the bubble is 0.365 m, resulting in a dimensionless distance of about 4.1 between the bubble and the two boundaries. The present theory exhibits good agreement with the EFEM results regarding the radius, period, and migration of the bubble at this distance parameter. While the EFEM-calculated pulsation pressure is slightly smaller than the theoretical prediction, this discrepancy can be attributed to numerical dissipation. Overall, our theory reproduces the time-dependent process of bubble pulsation pressure well compared with the traditional Keller equation.

\begin{figure*}
	\centering\includegraphics[width=0.9\linewidth]{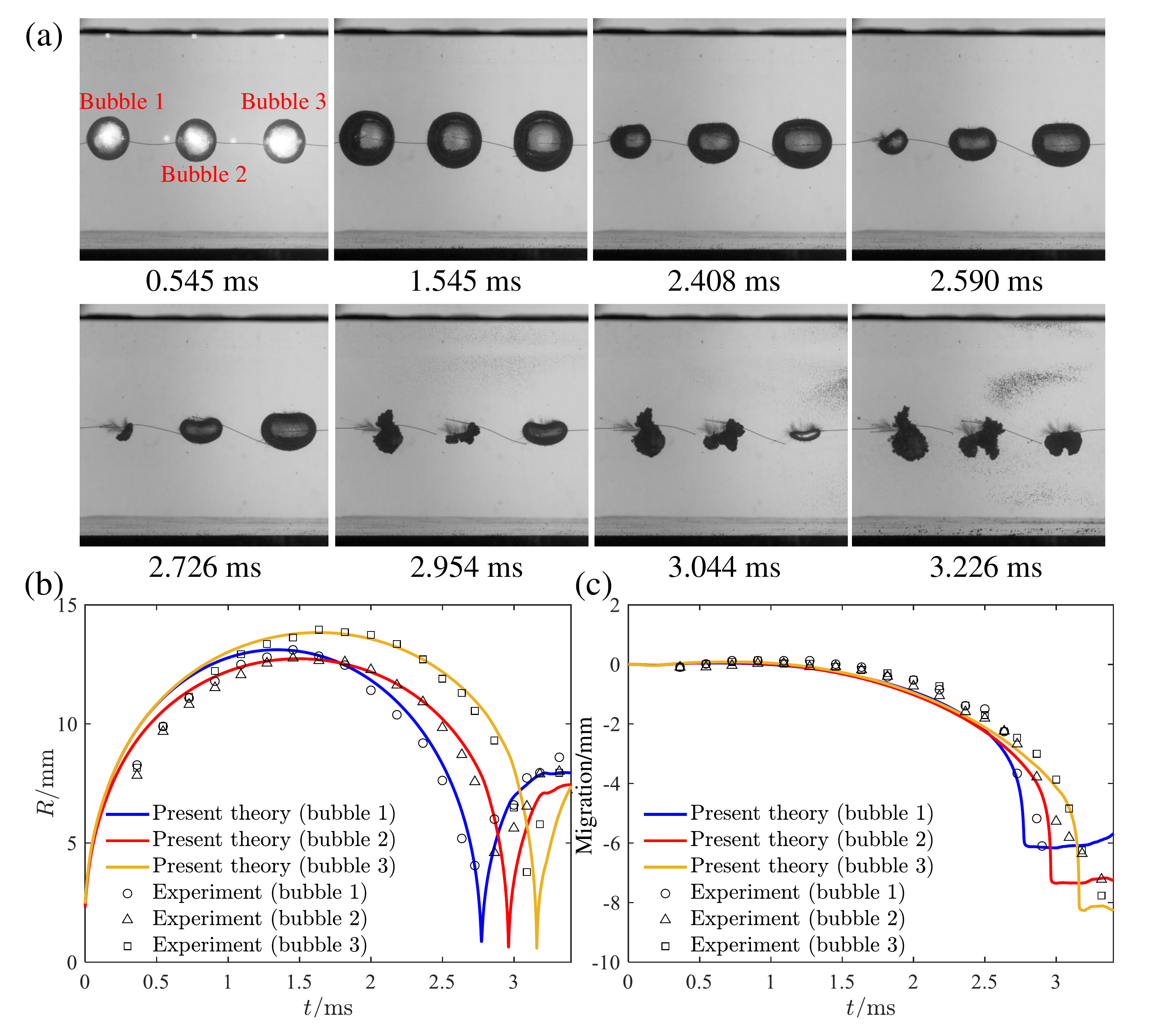}
	\caption{Comparison of the interaction of the three spark-induced bubbles with the theoretical results near the vertical sidewall, bottom and free surface. The effect of the sidewall is included in six horizontally equidistantly distributed bubbles induced by an identical voltage in the experiment. (a) Selected experimental images at typical moments. The sidewall is 20 mm from bubble 3. The three bubbles are initially located 49 mm from the free surface and 45 mm from the bottom. The distance between bubble 2 and bubble 1, as well as bubble 3, is 40 mm. Frame width: 119 mm. (b) Bubble radius (in theoretical calculations: $ R_{01} = 2.36 $ mm, $ R_{02} = 2.25 $ mm, $ R_{03} = 2.41 $ mm, $P_{01} =P_{02} =P_{03} =1.2$ MPa and ${\dot{R}}_{01} ={\dot{R}}_{02} ={\dot{R}}_{03} =$100 m/s). (c) Vertical bubble migration.}
	\label{mixthree}
\end{figure*}

Finally, we calculate the dynamics of three bubbles under the combined action of three boundaries, including the sidewall, bottom, and free surface, and compare the experimental results with the theoretical values, as shown in Fig. \ref{mixthree}. In the experiment, to equate the effect of the sidewall, six spark-induced bubbles are generated simultaneously by an identical voltage, and they are horizontally spaced at a distance of 40 mm. The three bubbles on the left are marked as bubble 1, bubble 2, and bubble 3, as illustrated in Fig. \ref{mixthree}(a). These configurations correspond to a sidewall on the right located 20 mm to bubble 3. All bubbles are initially positioned 49 mm from the free surface and 45 mm from the bottom. Due to the relatively free flow of fluids on the left side of bubble 1, bubble 1 is the first to collapse to its minimum volume. The pulsation of bubble 3 has the maximum pulsation period due to the influence of the right sidewall and two bubbles on the left. Throughout the entire pulsation process, all bubbles migrate downward in the vertical direction because the buoyancy effect of the bubbles is too weak to counteract the influence of the free surface and bottom. The initial radius for the bubbles in theory are: $ R_{01} = 2.36 $ mm, $ R_{02} = 2.25 $ mm and $ R_{03} = 2.41 $ mm. The initial inner bubble pressure and pulsation velocity of the three bubbles are 1.2 MPa and 100 m/s, respectively. The evolutions of the radius and migration of bubbles in the experiment agree well with the computed results in the present theory, as illustrated in Fig. \ref{mixthree}(b) and (c).

\subsection{Multiple bubble dynamics near boundary}\label{sect:discussion}

\begin{figure*}
	\centering\includegraphics[width=0.9\linewidth]{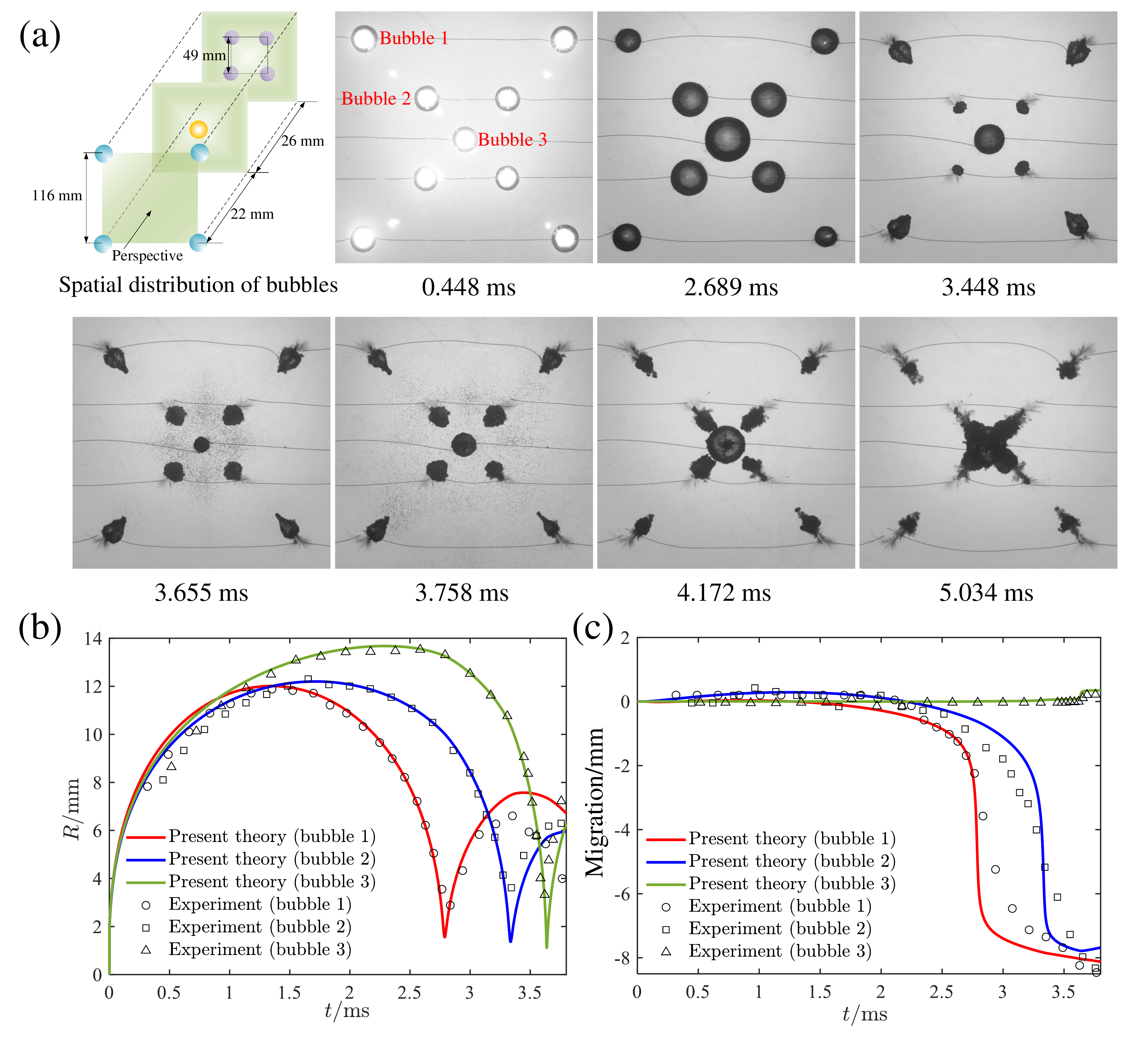}
	\caption{Comparison of the interaction of nine spark-induced bubbles with the theoretical results. (a) Selected experimental images at typical moments. The relative positions of nine bubbles are clarified by a sketch in the beginning. Frame width: 119 mm. (b) Bubble radius (in theoretical calculations: $ R_{01} =R_{03} = 1.94 $ mm, $ R_{02} = 1.88 $ mm, $P_{01} =P_{02} =P_{03} =1.2$ MPa and ${\dot{R}}_{01} ={\dot{R}}_{02} ={\dot{R}}_{03} =$120 m/s). (c) Vertical bubble migration.}
	\label{9bubble}
\end{figure*}

We first conducted an experiment with nine spark-induced bubbles distributed in the three-dimensional space and compared the obtained experimental values with theoretical results, as shown in Figure \ref{9bubble}. Fig. \ref{9bubble}(a) illustrates the relative positions of nine bubbles on the plane captured using a high-speed camera, where bubbles 1 and 2 are marked as representative bubbles of the outermost and second circles, respectively. Bubble 3 is identified as the central bubble. The nine bubbles are distributed in three planes. The outermost cycle of bubbles is situated in the closest plane to the field of view, whereas bubbles 2 and 3 are generated 48 mm and 22 mm away from the plane of bubble 1, respectively. Both the outermost and second circles of bubble array comprise four uniformly distributed bubbles on a $2 \times 2$ grid, as illustrated in Fig. \ref{9bubble}(a). The first circle's grid interval is set at 116 mm, while that of the second circle is 49 mm. As the positions of the nine bubbles are nearly symmetrical in the photographed plane, all the bubbles migrate toward the plane center. Note that some black dots are presented in the fourth frame, which are cavitation bubbles induced by pressure variation due to the rapid collapses and rebounds of bubbles. The outer bubble has a smaller pulsation period than the inner bubble due to the relatively free flow of fluid on the side away from the inner bubble. All bubbles have the same initial inner pressure and pulsation velocity in theory: $P_0 = 1.2$ MPa and $ \dot{R}_0= 120$ m/s. The initial radius of bubbles are: $ R_{01} =R_{03} = 1.94 $ mm and $ R_{02} = 1.88 $ mm. Considering the slight differences in bubble dynamics during most of the first pulsation cycle, the other bubbles in the outermost and second cycles have the same initial conditions as bubble 1 and bubble 2, respectively. Figure \ref{9bubble}(b) and (c) presents a comparison of the time-history of the bubble radius and vertical migration between theory and experiment. The present theory reproduces the dynamic features of bubbles observed in the experiment well, including the  pulsation period of bubbles, accelerated contraction of bubble 3 at the end of collapse and bubble migration direction.

\begin{figure*}
	\centering\includegraphics[width=1.0\linewidth]{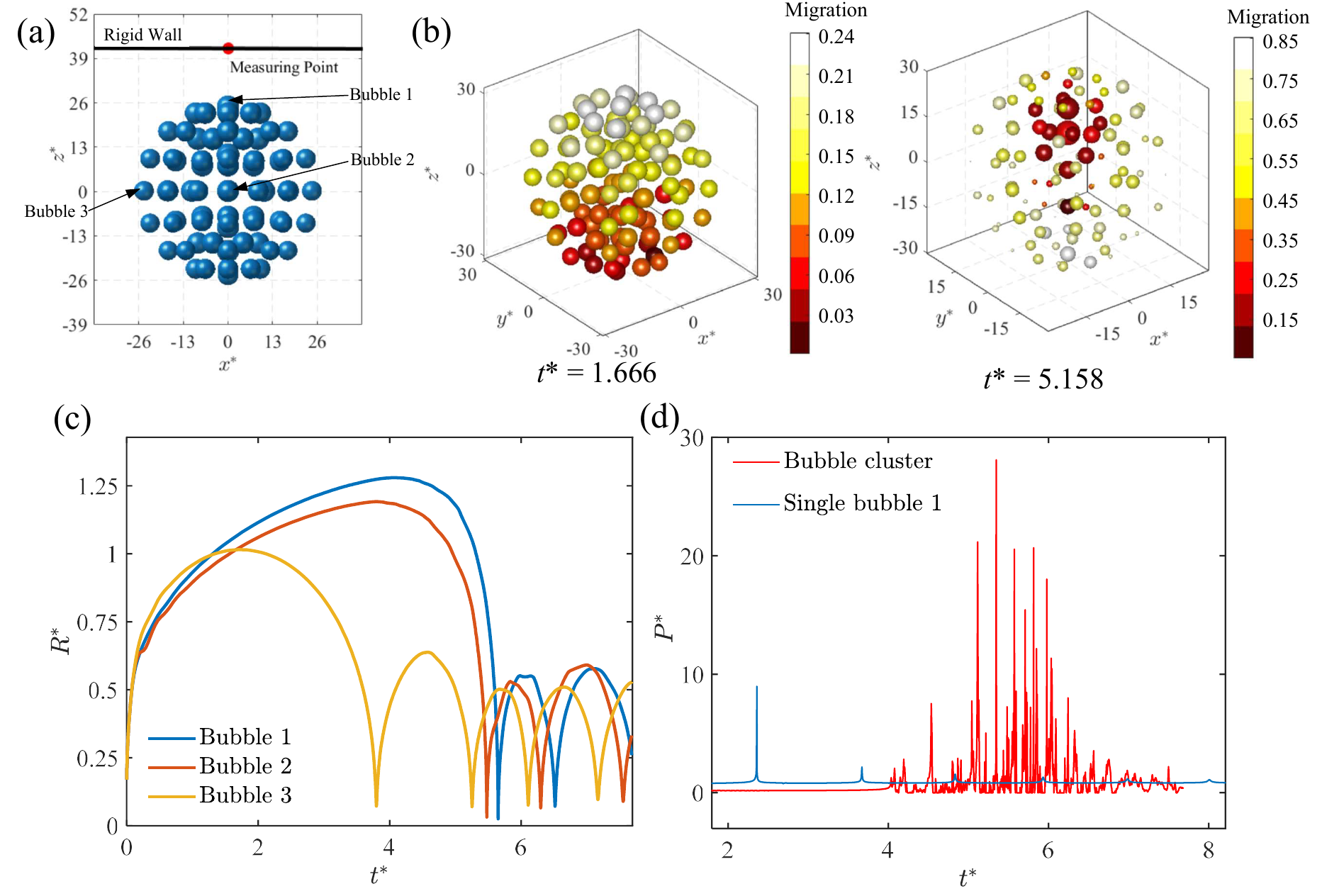}
	\caption{Computed dynamics of a spherical bubble cluster with 105 cavitation bubbles below a rigid wall. (a) Schematic diagram of the bubble cluster in the $xz$ plane. The black bold solid line denotes the rigid wall on which a pressure measurement point is placed directly above the bubble cluster. (b) The relative size and vertical migration of bubbles at two typical moments ($P_0^* = 11.55$, $ R_0^* = 0.13 $, $ \dot{R}_0^*= 17.65$). (c) Bubble radius for three typical bubbles. (d) Comparison of pressure at the measuring point between the bubble cluster and single bubble 1.}
	\label{array}
\end{figure*}

Next, as an application of the unified equation to the multiple bubbles, we discuss the dynamics of a spherical bubble cluster with 105 cavitation bubbles below a rigid wall. The bubble cluster comprises four layers of bubbles. The innermost layer has only one bubble located at the cluster center, while the outermost layer has 58 bubbles distributed evenly on a sphere with a radius of 195 mm. The radii of the middle two layers are 81 mm and 135 mm, respectively. The layer with a radius of 81 mm contains 14 bubbles uniformly distributed on it, and the layer with a radius of 135 mm contains 32 bubbles. The bubble cluster center is located 300 mm away from the wall. All bubbles have the same initial conditions: $P_0 = 1.2$ MPa, $ R_0 = 1.0 $ mm and $ \dot{R}_0= 180$ m/s. We designate three representative bubbles as bubbles 1, 2, and 3, as shown in Fig. \ref{array}(a). The maximum radius that a pulsating bubble can reach in a free field is defined as the reference length (7.6 mm). Fig. \ref{array}(b) presents the spatial distribution of bubbles at two typical moments in the dimensionless system, where the color denotes the amplitude of the vertical migration. At $t^*=1.666$, all bubbles are in their first pulsation cycle and show no significant migration tendency. As the distance from the wall increases, the migration tendency of the bubbles becomes weaker. However, at $t^*=4.598$, most of the bubbles, except for those still in the first pulsation cycle, migrate significantly toward the wall, resulting in the entire bubble cluster moving closer to the wall. This is reflected by the color bar values in Fig. \ref{array}(b). We also plot the radius of bubbles 1, 2, and 3 over time in Fig. \ref{array}(c) and the pressure evolution of the measuring point in Fig. \ref{array}(d). Bubble 3 has the shortest pulsation period, while bubble 1 has the longest, which is attributed to the difference in the number of surrounding bubbles, including real ones and mirror ones on the other side of the wall. During the collapse phase, the oscillation of bubble 1 accelerates because its surrounding bubbles rebound. These observations are similar to those of a planar bubble array observed in the previous study \cite{zhang2023unified}.
The bubble energy formula \cite{KhooWang2004} is employed to calculate the dimensionless bubble energy, and we found that the energy of bubble 1 in the first cycle increased from 4.19 when oscillating individually to 9.21 when in the bubble cluster. This indicates that some bubbles absorb the energy of other bubbles during bubble interaction. This also causes the wall pressure induced by some bubbles in the cluster to be more than two times higher than the pressure induced by bubble 1 alone. 	

\begin{figure*}
	\centering\includegraphics[width=1.0\linewidth]{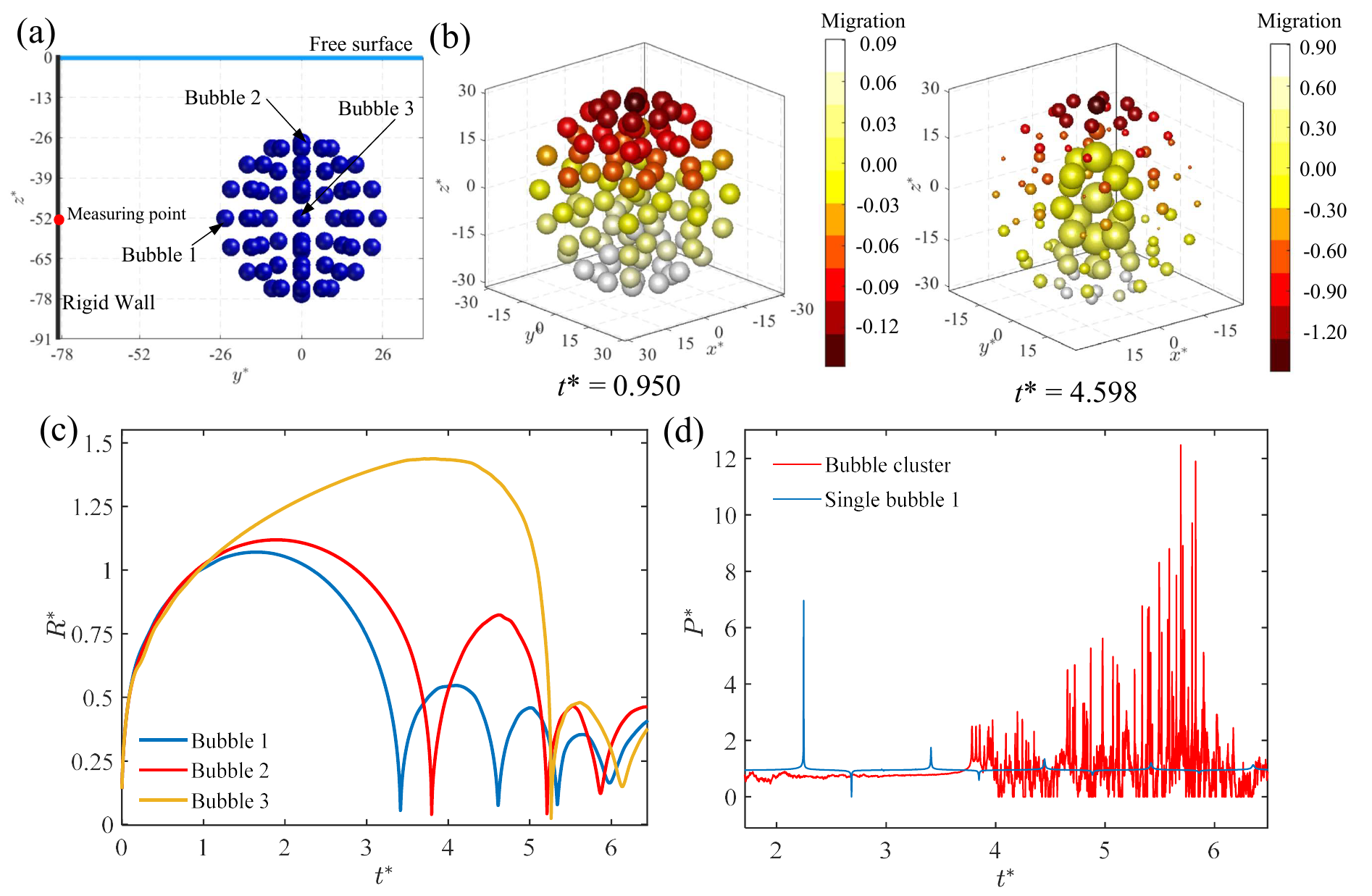}
	\caption{Computed dynamics of a spherical bubble cluster with 105 cavitation bubbles near a vertical rigid wall and a free surface. (a) Schematic diagram of the bubble cluster and boundary conditions in the $yz$ plane. The rigid wall is located at $y^*=79$ and the free surface is located at $z^*=0$. The bubble cluster center is at $x^*=y^*=0$ and $z^*=-52$. The pressure measurement point on the rigid wall is horizontal to the bubble cluster center. (b) Relative size and vertical migration of bubbles at two typical moments. The initial conditions for bubble pulsation are the same as those in Fig. \ref{array}. (c) Bubble radius for three typical bubbles. (d) Comparison of pressure at the measuring point between the bubble cluster and single bubble 1.}
	\label{mixb_cluster}
\end{figure*}

{\color{black}Finally, we compute the dynamics of the bubble cluster under the hybrid-boundary condition, comprising a free surface and a vertical rigid wall, as shown in Fig. \ref{mixb_cluster}. The initial conditions of the bubbles are the same as those in Fig. \ref{array}. Bubbles 1, 2, and 3 are marked in Fig. \ref{mixb_cluster}(a). The bubble cluster center is 52 from the free surface and 79 from the rigid wall in the dimensionless system. Fig. \ref{mixb_cluster}(b) illustrates the relative bubble size and vertical bubble migration at two typical moments. The vertical migration of bubbles gradually increases as the distance from the free surface increases. The bubbles at the top of the bubble cluster exhibit downward migration due to the influence of the free surface, while the bubbles at the bottom migrate upward due to the weaker effect of the free surface. At $t^*=0.950$, the bubbles are all in the initial expansion stage with relatively uniform sizes. However, at $t^*=4.598$, the bubbles inside the cluster are relatively larger, which can also be reflected in the radius-time curve of the bubbles in Fig. \ref{mixb_cluster}(c). The maximum radius and period of bubble 3 significantly exceed those of bubbles 1 and 2, as the pulsation of the central bubble is slowed down by the surrounding contracting bubbles at the late stage of the first cycle. 
In Fig. \ref{mixb_cluster}(d), we present the wall pressure induced by the bubble cluster and bubble 1 alone. The peak pressure on the wall does not increase as much as that in Fig. \ref{array} due to the effect of the free surface. Nevertheless, the wall pressure induced by the bubble cluster significantly exceeds that of the single bubble 1, with its pressure peak reaching close to twice that of single bubble 1.}

\section{Summarization and Conclusion}\label{sect:conclusion}
In this study, the unified equation for bubble dynamics \cite{zhang2023unified} was rederived using the moving point source and dipole. We found that the form of the bubble pulsation equation was the same as that reported by Zhang et al. \cite{zhang2023unified} considering the motion of singularities. The theory is applied to situations involving multiple bubbles and hybrid boundaries. The study's findings are summarized as follows:


\noindent1. The compressibility of bubble migration has an important effect on bubble dynamics behavior. The present theory fully considers the coupling between compressible bubble pulsation and migration in a compressible flow field.

\noindent2. A finite number of mirror bubbles are included to model the bubble dynamics near crossed boundaries. An infinite number of mirror bubbles must be introduced to account for the effect of parallel boundaries. Through calculations with various numbers of mirror bubbles for the parallel boundaries, all the results converge to within 0.1\% when the number of mirror bubbles on the side of a boundary reaches 30. A good achievement is obtained between the theoretical result and the experiment and simulation, which demonstrate the capability of the present theory to predict bubble dynamics near hybrid boundaries.  

\noindent3. A spherical bubble cluster containing more than one hundred cavitation bubbles is computed using the present theory under two boundary conditions: one is below a rigid wall and the other is near a vertical rigid wall below the free surface. The theoretical results indicated that some bubbles in the cluster could produce pressure peaks of more than double those of the individual bubble on the surrounding structures. 

This study is based on the assumption of spherical bubbles, which could be used to predict the main characteristics of bubbles, including bubble size, period, migration, and pulsation pressure. For strong asymmetrical deformations of bubbles, such as splitting, breaking into pieces, and dissolving into the liquid, advanced numerical and experimental methods are required to conduct in-depth research.

\Acknowledgements{This work is funded by the National Natural Science Foundation of China (52088102, 51925904), the National Key R\&D Program of China (2022YFC2803500, 2018YFC0308900), Finance Science and Technology Project of Hainan Province (ZDKJ2021020), and the Xplore Prize. }

\InterestConflict{The authors declare that they have no conflict of interest.}

\end{multicols}


\begin{thebibliography}{99}
	
	\bibitem{loewen1996bubbles}
	M. Loewen, M. O'Dor, and M. Skafel,
	\newblock { J. Geophys. Res.-Oceans} {\textbf{101}}, 20759--20769 (1996).
	
	\bibitem{Deike2022}
	L. Deike,
	\newblock { Annu. Rev. Fluid Mech.} {\textbf{54}}, 191--224 (2022).
	
	\bibitem{Watson19}
	L.~M. Watson, J.~Werpers, and E.~M. Dunham,
	\newblock { Geophys.} {\textbf{84}}, 27--45 (2019).
	
	\bibitem{Nourhani2020}
	A.~Nourhani, E.~Karshalev, F.~Soto, and J.~Wang,
	\newblock { Research} {\textbf{2020}}, 7823615 {(2020)}.
	
	\bibitem{Lohse2022}
	D.~Lohse,
	\newblock { Annu. Rev. Fluid Mech.} {\textbf{54}}, 349--382 {(2022)}.
	
	\bibitem{turkoz2018impulsively}
	E. Turkoz, A. Perazzo, H. Kim, H.~A Stone, and C. Arnold,
	\newblock {Phys. Rev. Lett.} {\textbf{120}}, 074501 (2018).
	
	\bibitem{ji2016numerical}
	B.~Ji, J.~Wang, X.~Luo, K.~Miyagawa, L.-Z Xiao, X.~Long, and Y.~Tsujimoto,
	\newblock { J. Mech. Sci. Technol.} {\textbf{30}}, 2507--2514 (2016).
	
	\bibitem{spzh14}
	A.~D. Stroock, V.~V. Pagay, M.~A. Zwieniecki, and N.~M. Holbrook,
	\newblock { Annu. Rev. Fluid Mech.} {\textbf{46}}, 615--642 (2014).
	
	\bibitem{carling2017bubble}
	P.~A. Carling, M.~Perillo, J.~Best, and M.~Garcia,
	\newblock { Earth Surf. Process. Landf.} {\textbf{42}}, 1308--1316 (2017).
	
	\bibitem{Yang2022}
	Y.~Tao, P.~Wu, Y.~Dai, X.~Luo, S.~Manickam, D.~Li, Y.~Han, and P.~Loke~Show,
	\newblock { Chem. Eng. J.} {\textbf{436}}, 135158 (2022).
	
	\bibitem{sc19}
	E.~Stride and C.~Coussios,
	\newblock { Nat. Rev. Phys.} {\textbf{1}}, 495--509 (2019).
	
	\bibitem{Dollet2019}
	B.~Dollet, P.~Marmottant, and V.~Garbin,
	\newblock { Annu. Rev. Fluid Mech.} {\textbf{51}}, 331--355 (2019).
	
	\bibitem{Rod2015}
	J.~Rodrguez, A.~Sevilla, C.~Martnez-Bazn, and J.~Gordillo,
	\newblock { Annu. Rev. Fluid Mech.} {\textbf{47}}, 405--429 (2015).
	
	\bibitem{Mukundakrishnan2009}
	K.~Mukundakrishnan, P.~Ayyaswamy, and D.~Eckmann,
	\newblock {\em J. Biomech Eng.} {\textbf{131}}, 074516 (2009).
	
	\bibitem{vas12}
	V.~B. Veljkovic, J.~M. Avramovic, and O.~S. Stamenkovic,
	\newblock { Renew. Sust. Energ. Rev.} {\textbf{16}}, 1193--1209 (2012).
	
	\bibitem{Lohse2018D}
	D.~Lohse,
	\newblock { Phys. Rev. Fluids} {\textbf{3}}, 110504 (2018).
	
	\bibitem{vshl00}
	M.~Versluis, B.~Schmitz, A.~von~der Heydt, and D.~Lohse,
	\newblock { Science}. {\textbf{289}}, 2114--2117 (2000).
	
	\bibitem{Cole1948book}
	R.~H. Cole,
	\newblock in {\em Underwater Explosions},
	\newblock edited by Dover (Princeton University Press, America, 1948), pp. 3-433.
	
	\bibitem{Cui2016}
	P.~Cui, A.-M Zhang, and S.-P Wang,
	\newblock { Phys. Fluids} {\textbf{28}}, 117103 (2016).
	
	\bibitem{zhang2021engineering}
	J.~Zhang, S.~Wang, X.~Jia, Y.~Gao, and F.~Ma,
	\newblock { Phys. Fluids} {\textbf{33}}, 017118 (2021).
	
	\bibitem{lhfww19}
	J.~Lyons, M.~Haney, D.~Fee, A.~Wech, and C.~Waythomas,
	\newblock { Nat. Geosci.} {\textbf{12}}, 952--958 (2019).
	
	\bibitem{Ziolkowski1982}
	A.~Ziolkowski, G.~Parkes, L.~Hatton, and T.~Haugland,
	\newblock { Geophys.} {\textbf{47}}, 1413--1421 (1982).
	
	\bibitem{de2014bubble}
	K.~L. de~Graaf, P.~A. Brandner, and I.~Penesis,
	\newblock {Exp. Therm. Fluid Sci.} {\textbf{55}}, 228--238 (2014).
	
	\bibitem{wehner2020acoustic}
	D.~Wehner, U.~P. Svensson, and M.~Landr{\o},
	\newblock {J. Acoust. Soc. Am.} {\textbf{147}}, 1092--1103 (2020).
	
	\bibitem{Aristoff2009}
	J.~M. Aristoff and J.~W.~M. Bush,
	\newblock { J. Fluid. Mech.} {\textbf{619}}, 45--78 (2009).
	
	\bibitem{Truscott2014}
	T.~Truscott, B.~Epps, and J.~Belden,
	\newblock { J. Fluid Mech.} {\textbf{46}}, 355--378 (2014).
	
	\bibitem{Wentao2023}
	W.-T Liu, A.-M Zhang, X.-H Miao, F.-R Ming, and Y.-L Liu,
	\newblock { J. Fluid Mech.} {\textbf{958}}, A42 (2023).
	
	\bibitem{Brennen2015}
	C.~E. Brennen,
	\newblock {Interface Focus}, {\textbf{5}}, 20150022 (2015).
	
	\bibitem{lhgh18}
	Y.~Liu, D.~He, X.~Gong, and H.~Huang,
	\newblock { J. Fluid Mech.} {\textbf{844}}, 567--596 (2018).
	
	\bibitem{omata2022ultrasound}
	D.~Omata, L.~Munakata, S.~Kageyama, Y.~Suzuki, T.~Maruyama, T.~Shima,
	T.~Chikaarashi, N.~Kajita, K.~Masuda, N.~Tsuchiya, K.~Maruyama, and
	R.~Suzuki,
	\newblock { J. Drug Target.} {\textbf{30}}, 200--207 (2022).
	
	\bibitem{r17}
	{Lord Rayleigh},
	\newblock { Philos. Mag.} {\textbf{34}}, 94--98 (1917).
	
	\bibitem{plesset49}
	M.~Plesset,
	\newblock { J. Appl. Mech.} {\textbf{16}}, 277--282 (1949).
	
	\bibitem{h70}
	A.~N. Hicks,
	\newblock {\em Effect of bubble migration on explosion-induced whipping in ships},
	\newblock Technical Report (Naval Ship Research and Development Center,
	Bethesda, MD, 1970).
	
	\bibitem{km80}
	J.~Keller and M.~Miksis,
	\newblock { J. Acoust. Soc. Am.} {\textbf{68}}, 628 (1980).
	
	\bibitem{Chahine2009}
	G.~Chahine,
	\newblock { J. Hydrodyn.} {\textbf{21}}, 316--332 (2009).
	
	\bibitem{zhang2023unified}
	A-.M Zhang, S.-M Li, P.~Cui, S. Li, and Y.-L Liu.
	\newblock { Phys. Fluids} {\textbf{35}}, 033323 (2023).
	
	\bibitem{Gonzalez2021}
	S.~R. Gonzalez-Avila, F.~Denner, and C.~D. Ohl,
	\newblock { Phys. Fluids} {\textbf{33}}, 032118 (2021).
	
	\bibitem{Blakewall1986}
	J.~R. Blake, B. Taib, and G. Doherty,
	\newblock {J. Fluid Mech.} {\textbf{170}}, 479--497 (1986).
	
	\bibitem{qianxi2014}
	Q.~Wang,
	\newblock { J. Fluid Mech.} {\textbf{745}}, 509--536 (2014).
	
	\bibitem{saade2021}
	Y.~Saade, M.~Jalaal, A.~Prosperetti, and D.~Lohse,
	\newblock { J. Fluid Mech.} {\textbf{926}}, A5 (2021).
	
	\bibitem{Koukouvinis2016}
	P.~Koukouvinis, M.~Gavaises, O.~Supponen, and M.~Farhat,
	\newblock {\em Phys. Fluids} {\textbf{28}}, 052103 (2016).
	
	\bibitem{Goh2017}
	B.~Goh, S.~Gong, S.-W Ohl, and B.~C Khoo,
	\newblock { Int. J. Multiph. Flow} {\textbf{90}}, 156--166 (2017).
	
	\bibitem{tokk06}
	C.~K. Turangan, G.~P. Ong, E.~Klaseboer, and B.~C. Khoo,
	\newblock { J. Appl. Phys.} {\textbf{100}}, 054910 (2006).
	
	\bibitem{Brujan2001}
	E.-A Brujan, K.~Nahen, P.~Schmidt, and A.~Vogel,
	\newblock { J. Fluid Mech.} {\textbf{433}}, 283--314 (2001).
	
	\bibitem{Han2015}
	R.~Han, A-.M Zhang, and Y.-L Liu,
	\newblock { Ocean Eng.} {\textbf{110}}, 325--338 (2015).
	
	\bibitem{Tong2019}
	T.~Li, A.-M Zhang, S.-P Wang, G.-Q Chen, and S.~Li,
	\newblock { Phys. Fluids} {\textbf{31}}, 092108 (2019).
	
	\bibitem{liang2021interaction}
	W.~Liang, R.~Chen, J.~Zheng, X.~Li, and F.~Lu,
	\newblock {\em Phys. Fluids} {\textbf{33}}, 067107 (2021).
	
	\bibitem{Tomita2017s}
	Y.~Tomita and K.~Sato,
	\newblock { J. Fluid Mech.} {\textbf{819}}, 465--493 (2017).
	
	\bibitem{liu2021interaction}
	N.-N Liu, A.-M Zhang, P.~Cui, S.-P Wang, and S.~Li,
	\newblock {\em Phys. Fluids} {\textbf{33}}, 106103 (2021).
	
	\bibitem{Han2019s}
	R.~Han, L.~Tao, A.-M Zhang, and S.~Li,
	\newblock { Ocean Eng.} {\textbf{31}}, 062107 (2019).
	
	\bibitem{Chen2022}
	R.~Chen, W.~Liang, J.~Zheng, X.~Li, and Y.~Lin,
	\newblock { Phys. Fluids} {\textbf{34}}, 037105 (2022).
	
	\bibitem{Zwaan2007}
	Z.~Ed, S.~L. Gac, T.~Kinko, and C.-D Ohl,
	\newblock {\em Phys. Rev. Lett.} {\textbf{98}}, 254501 (2007).
	
	\bibitem{amzhang2023}
	A.-M Zhang, S.-M Li, P.~Cui, S.~Li, and Y.-L Liu,
	\newblock {\em Theor. App. Mech. Lett.} {\textbf{13}}, 100438 (2023).
	
	\bibitem{zhang2019numerical}
	L. Zhang, J. Zhang, and J. Deng,
	\newblock {\em Phys. Rev. E} {\textbf{99}}, 043108 (2019).
	
	\bibitem{Doinikov07}
	A.~Doinikov,
	\newblock { J. Acoust. Soc. Am.} {\textbf{116}}, 821 (2007).
	
	\bibitem{Maeda2019}
	K.~Maeda and T.~Colonius,
	\newblock { J. Fluid Mech.} {\textbf{862}}, 1105--1134 (2019).
	
	\bibitem{RN213}
	S.-M Li, P~Cui, S~Zhang, W.-T Liu, and Y.-X Peng,
	\newblock { China Ocean Eng.} {\textbf{34}}, 828--839 (2020).
	
	\bibitem{Qianxi20c}
	Q.~Wang, M.~Mahmud, J.~Cui, W.~R. Smith, and A.~D. Walmsley,
	\newblock { Phys. Fluids} {\textbf{32}}, 053306 (2020).
	
	\bibitem{CuiRN2020}
	J.~Cui, Z.-P Chen, Q.~Wang, T.-R Zhou, and C.~Corbett,
	\newblock { Ultrason. Sonochem.} {\textbf{64}}, 104951 (2020).
	
	\bibitem{Shimin21c}
	S.-M Li, Y.-L Liu, Q.~Wang, and A.-M Zhang,
	\newblock { Phys. Fluids} {\textbf{33}}, 073310 (2021).
	
	\bibitem{Yun2016c}
	Y.-L Liu, Q.~Wang, S.-P Wang, and A.-M Zhang,
	\newblock { Phys. Fluids} {\textbf{28}}, 122101 (2016).
	
	\bibitem{ZhangS2017c}
	S.~Zhang, A.-M Zhang, S.-P Wang, and J.~Cui,
	\newblock { Phys. Fluids} {\textbf{29}}, 092107 (2017).
	
	\bibitem{Kiyama2021}
	A.~Kiyama, T.~Shimazaki, J.~Gordillo, and Y.~Tagawa,
	\newblock { Phys. Rev. Fluids} {\textbf{6}}, 083601 (2021).
	
	\bibitem{Shim2023v}
	S.-M Li, A.-M Zhang, P~Cui, S~Li, and Y.-L Liu,
	\newblock { J. Fluid Mech.} {\textbf{962}}, A28 (2023).
	
	\bibitem{shimin2019}
	S.-M Li, A.-M Zhang, Q.~Wang, and S.~Zhang,
	\newblock { Phys. Fluids} {\textbf{31}}, 107105 (2019).
	
	\bibitem{Blake87b}
	J.~R. Blake and D.~C. Gibson,
	\newblock { Annu. Rev. Fluid Mech.} {\textbf{19}}, 99--123 (1987).
	
	\bibitem{Zhangm2013}
	A.-M Zhang and B.-Y Ni,
	\newblock { Sci. China-Phys. Mech. Astron.} {\textbf{56}}, 2162--2169 (2013).
	
	\bibitem{Wang03BIM}
	C~Wang, B.~C. Khoo, and K.~S. Yeo,
	\newblock { Comput. Fluids} {\textbf{32}}, 1195--1212 (2003).
	
	\bibitem{Hanrui2021}
	R.~Han, A.-M Zhang, S.~Tan, and S.~Li,
	\newblock { J. Fluid Mech.} {\textbf{932}}, A8 (2022).
	
	\bibitem{Liu19i}
	Y.-L Liu, A.-M Zhang, Z.-L Tian, and S.-P Wang,
	\newblock { Phys. Fluids} {\textbf{31}}, 092111 (2019).
	
	\bibitem{LiuNN2020}
	N.-N Liu, A.-M Zhang, Y.-L Liu, and T. Li.
	\newblock { Phys. Fluids} {\textbf{32}}, 046107 (2020).
	
	\bibitem{Wang16d}
	L.-K Wang, Z.-F Zhang, and S.-P Wang,
	\newblock { Appl. Ocean Res.} {\textbf{59}}, 183--192 (2016).
	
	\bibitem{Zeng20c}
	Q.~Zeng, S. R.~Gonzalez, and C.-D Ohl,
	\newblock { J. Fluid Mech.} {\textbf{896}}, A28 (2020).
	
	\bibitem{Zeng18c}
	Q.~Zeng, S.~R. Gonzalez-Avila, R.~Dijkink, P.~Koukouvinis, M.~Gavaises, and
	C.-D Ohl,
	\newblock { J. Fluid Mech.} {\textbf{846}}, 341--355 (2018).
	
	\bibitem{Lechner20c}
	C.~Lechner, W.~Lauterborn, M.~Koch, and R.~Mettin,
	\newblock { Phys. Rev. Fluids} {\textbf{5}}, 093604 (2020).
	
	\bibitem{Zhang15ph}
	A-.M Zhang, P.~Sun, and F.~Ming,
	\newblock { Comput. Methods Appl. Mech. Engrg.} {\textbf{294}}, 189--209 (2015).
	
	\bibitem{li2022improved}
	M.-K Li, A.-M Zhang, Y.-X Peng, and F.-R Ming,
	\newblock { J. Comput. Phys.} {\textbf{458}}, 111106 (2022).
	
	\bibitem{RN112021}
	M.~Vassholz, H.~P. Hoeppe, J.~Hagemann, J.~M. Rosselló, M.~Osterhoff,
	R.~Mettin, T.~Kurz, A.~Schropp, F.~Seiboth, C.~G. Schroer, M.~Scholz,
	J.~Möller, J.~Hallmann, U.~Boesenberg, C.~Kim, A.~Zozulya, W.~Lu,
	R.~Shayduk, R.~Schaffer, A.~Madsen, and T.~Salditt,
	\newblock {\em Nat. Commun.} {\textbf{12}}, 3468 (2021).
	
	\bibitem{altmvl01}
	I.~Akhatov, O.~Lindau, A.~Topolnikov, R.~Mettin, N.~Vakhitova, and
	W.~Lauterborn,
	\newblock {\em Phys. Fluids} {\textbf{13}}, 2805 (2001).
	
	\bibitem{ro21}
	J.~Rossello and C.-D. Ohl,
	\newblock { Phys. Rev. Lett.} {\textbf{127}}, 044502 (2021).
	
	\bibitem{Jingzhu2021}
	J.~Wang, H.~Li, W.~Guo, Z.~Wang, T.~Du, Y.~Wang, A.~Abe, and C.~Huang,
	\newblock { J. Fluid Mech.} {\textbf{919}}, A42 (2021).
	
	\bibitem{Dadvand2012}
	A.~Dadvand, B.~C. Khoo, M.~Shervani, and S.~Khalilpourazary,
	\newblock { Eng. Anal. Bound. Elem.} {\textbf{36}}, 1595--1603 (2012).
	
	\bibitem{okdbnf06}
	D.~Obreschkow, P.~Kobel, N.~Dorsaz, A.~de~Bosset, C.~Nicollier, and M.~Farhat,
	\newblock { Phys. Rev. Lett.} {\textbf{97}}, 094502 (2006).
	
	\bibitem{sun2022cavitation}
	Y. Sun, Z. Yao, H. Wen, Q. Zhong, and F. Wang,
	\newblock { Phys. Fluids} {\textbf{34}}, 073314 (2022).
	
	\bibitem{hh10}
	C.~F. Hung and J.~J. Hwangfu,
	\newblock { J. Fluid Mech.} {\textbf{651}}, 55--80 (2010).
	
	\bibitem{Shimin21d}
	S.-M Li, A.-M Zhang, and N.-N Liu,
	\newblock { Theor. App. Mech. Lett.} {\textbf{11}}, 100311 (2021).
	
	\bibitem{Lee2007}
	M.~Lee, E.~Klaseboer, and B.~C. Khoo,
	\newblock { Journal of Fluid Mechanics} {\textbf{570}}, 407--429 (2007).
	
	\bibitem{wb10}
	Q.~Wang and J. R.~Blake,
	\newblock { J. Fluid Mech.} {\textbf{659}}, 191--224 (2010).
	
	\bibitem{Ma18J}
	J.~Ma, C.~T. Hsiao, and G.~L. Chahine,
	\newblock { Ultrason. Sonochem.} {\textbf{40}}, 944--954 (2018).
	
	\bibitem{Best1991}
	J.~Best,
	\newblock {\em The dynamics of underwater explosions},
	\newblock Dissertation for the Doctoral Degree (University of Wollongong, Wollongong, 1991), pp. 1-257.
	
	\bibitem{Tian2020OE}
	Z.-L Tian, Y.-L Liu, A.-M Zhang, and L. Tao,
	\newblock { Ocean Eng.} {\textbf{196}}, 106714 (2020).
	
	\bibitem{w13}
	Q.~Wang,
	\newblock { Phys. Fluids} {\textbf{25}}, 072104 (2013).
	
	\bibitem{KhooWang2004}
	C.~Wang and B.~C. Khoo,
	\newblock {J. Comput. Phys.} {\textbf{194}}, 451--480 (2004).
	
\end{thebibliography}
\end{document}